%% file: mgirardi.tex
\documentclass[bibyear]{aa}
\usepackage{graphicx}
\usepackage{txfonts}
\newcommand{\mincir}{\raise -2.truept\hbox{\rlap{\hbox{$\sim$}}\raise5.truept
\hbox{$<$}\ }}
\newcommand{\magcir}{\raise -2.truept\hbox{\rlap{\hbox{$\sim$}}\raise5.truept
\hbox{$>$}\ }}
\newcommand{\siml}{\raise -2.truept\hbox{\rlap{\hbox{$\sim$}}\raise5.truept
\hbox{$<$}\ }}
\newcommand{\simg}{\raise -2.truept\hbox{\rlap{\hbox{$\sim$}}\raise5.truept
\hbox{$>$}\ }}
\newcommand{\be}{\begin{equation}}
\newcommand{\ee}{\end{equation}}
\newcommand{\ba}{\begin{eqnarray}}
\newcommand{\ea}{\end{eqnarray}}
\newcommand {\kpc} {$h_{70}^{-1}$ kpc $\;$}

\newcommand {\hh} {$h_{70}^{-1}$ Mpc}

\newcommand {\kss} {km~s$^{-1}$}

\newcommand {\mquii} {$\times 10^{15}\;h_{70}^{-1}\;M_{\odot}$}

\newcommand{\arcm}{\ensuremath{\mathrm{^\prime}\;}}

\begin{document}
\title{Fossil groups origins \\ III. The relation between optical and X-ray luminosities}

   \author{
M. Girardi\inst{1,2}
\and J. A. L. Aguerri\inst{3,4}
\and S. De Grandi\inst{5}
\and E. D'Onghia\inst{6,7}
\and R. Barrena\inst{3,4}
\and W. Boschin\inst{8}
\and J. M\'endez-Abreu\inst{3,4,9}
\and R. S\'anchez-Janssen\inst{10}
\and S. Zarattini\inst{3,4}
\and A. Biviano\inst{2}
\and N. Castro-Rodriguez\inst{3,4}
\and E. M. Corsini\inst{11,12}
\and C. del Burgo\inst{13}
\and J. Iglesias-P\'aramo\inst{14,15}
\and J. M. Vilchez\inst{14}
}

   \offprints{M. Girardi, \email{girardi@oats.inaf.it}}

   \institute{ 
     Dipartimento di Fisica dell'Universit\`a degli Studi
     di Trieste - Sezione di Astronomia, via Tiepolo 11, I-34143
     Trieste, Italy 
\and INAF - Osservatorio Astronomico di Trieste,
     via Tiepolo 11, I-34143 Trieste, Italy 
\and Instituto de
     Astrof\'{\i}sica de Canarias, C/V\'{\i}a L\'actea s/n, E-38205 La
     Laguna (Tenerife), Canary Islands, Spain
\and Departamento de
     Astrof\'{\i}sica, Universidad de La Laguna, Av. del
     Astrof\'{\i}sico Franciso S\'anchez s/n, E-38205 La Laguna
     (Tenerife), Canary Islands, Spain
\and 
     INAF -- Osservatorio Astronomico di Brera, via E. Bianchi 46,
     I-23807 Merate (LC), Italy
\and Astronomy Department, University of Wisconsin, 475 Charter St., Madison, WI 53706, USA
\and Alfred P. Sloan Fellow
\and 
     Fundaci\'on Galileo
     Galilei -- INAF, Rambla Jos\'e Ana Fern\'andez Perez 7, E-38712
     Bre\~na Baja (La Palma), Canary Islands, Spain
\and School of Physics and Astronomy, University of St Andrews, North Haugh, St Andrews, KY16 9SS, U.K. (SUPA)
\and NRC Herzberg Institute of Astrophysics, 5071 West Saanich Road, Victoria, BC, V9E 2E7, Canada
\and Dipartimento di Fisica e Astronomia `G. Galilei', Universit\`a di Padova, vicolo dell'Osservatorio 3, 35122 Padova, Italy 
\and INAF -- Osservatorio Astronomico di Padova, vicolo dell'Osservatorio 5, I-35122 Padova, Italy 
\and Instituto Nacional de Astrof\'isica, \'Optica y Electr\'onica (INAOE), Aptdo. Postal 51 y 216, 72000 Puebla, Pue., Mexico
\and Instituto de Astrof\'isica de Andalucia--C.S.I.C., E-18008 Granada, Spain 
\and Centro Astron\'omico Hispano Alem\'an, C/ Jes\'us Durb\'an Rem\'on 2-2. 04004, Almer\'ia, Spain 
}

\date{Received  / Accepted }

\abstract {} {This study is part of the FOssil Groups Origin (FOGO)
  project which aims to carry out a systematic and multiwavelength
  study of a large sample of fossil systems. Here we focus on the
  relation between the optical luminosity ($L_{\rm opt}$) and X-ray
  luminosity ($L_{\rm X}$).} {Out of a total sample of 28 candidate fossil
  systems, we consider a sample of 12 systems whose fossil
  classification has been confirmed by a companion study. They are
  compared with the complementary sample of 16 systems whose
      fossil nature has  not been confirmed  and with a subsample of 102
  galaxy systems from the RASS-SDSS galaxy cluster survey.  Fossil and
  normal systems span the same redshift range $0<z<0.5$ and have the
  same $L_{\rm X}$ distribution. For each fossil system, the $L_{\rm
    X}$ in the 0.1--2.4\,keV band is computed using data from the
  {{\em ROSAT}} All Sky Survey  to be comparable to the estimates of
  the comparison sample. For each fossil and normal system we
  homogeneously compute $L_{\rm opt}$ in the $r$-band within the 
    characteristic cluster radius, using data from the Sloan Digital
  Sky Survey Data Release 7.}  {We sample the $L_{\rm X}$--$L_{\rm
    opt}$ relation over two orders of magnitude in $L_{\rm X}$.  Our
  analysis shows that fossil systems are not statistically
  distinguishable from the normal systems through the 2D
  Kolmogorov-Smirnov test nor the fit of the $L_{\rm X}$--$L_{\rm
    opt}$ relation. Thus, the optical luminosity of the galaxy system
  does strongly correlate with the X-ray luminosity of the hot gas
  component, independently of whether the system is fossil or
  not.  We discuss our results in comparison with previous
  literature.}  {We conclude that our results are consistent with the
  classical merging scenario of the brightest galaxy formed via
  merger/cannibalism of other group galaxies with conservation of the
  optical light. We find no evidence for a peculiar state of the hot
  intracluster medium.}


\titlerunning{Fossil Groups Origins. III.} 

\maketitle
%
%________________________________________________________________

\section{Introduction}
\label{intro}

Several studies of galaxy systems have revealed an interesting class
of objects termed fossil groups (Ponman et al. \cite{pon94}).  From
the observational point of view, these are defined as galaxy systems
with a magnitude difference of at least two magnitudes---in the
$R$-band---between the brightest group/cluster galaxy (BCG) and the
second-brightest galaxy within half the virial radius
$R_{200}$\footnote{The radius $R_{\delta}$ is the radius of a sphere
  with mass overdensity $\delta$ times the critical density at the
  redshift of the galaxy system.} and an extended thermal X-ray halo
with bolometric X-ray luminosity $L_{\rm X}{\rm (bol)}> 10^{42}\ h_{50
}^{-2}$ erg s $^{-1}$ (see Jones et al. \cite{jon03} for the
rationale).  Thus, the fossil groups appear to be extreme environments
devoid of typical bright galaxies while simultaneously being home to
the brightest and most massive galaxies in the Universe.  The first
explanation was that they are old, isolated galaxy systems in
which the large galaxies have merged or coalesced through dynamical
friction. In this merging scenario, the magnitude gap shown by the
fossil systems is a consequence of evolution rather than an initial deficit
of $\sim L^*$ galaxies (i.e., the failed group scenario; see, e.g.,
the discussion in the study of Mulchaey \& Zabludoff \cite{mul99}).

The merging scenario has been invoked to explain such observational
features as the high values of X-ray luminosity ($L_{\rm X}$) and
temperature ($T_{\rm X}$) of fossil systems with respect to those of
normal systems with comparable optical luminosity ($L_{\rm opt}$) or
comparable velocity dispersion ($\sigma_{v}$; six fossil groups in
Jones et al. \cite{jon03}; seven in Khosroshahi et
al. \cite{kho07}) and some evidence of a high centrally
concentrated dark matter halo (Khosroshahi et al. \cite{kho06}).  The
above differences with normal systems have been generally interpreted
as due to an early formation epoch of fossil groups as suggested by
numerical simulations (e.g., D'Onghia et
al. \cite{don05}). Accordingly, the BCGs of fossil groups should contain a
fossil relic of the structure formation in the high-redshift
Universe. Early observations have revealed that the BCGs of fossil
groups have different observational properties than other bright
elliptical galaxies, their discy isophotes (seven fossil groups;
Khosroshahi et al. \cite{kho06}), for example, supporting the idea that they are
formed from gas-rich mergers in early times.

More recent studies have opened the discussion about the special
nature of fossil groups.  Alternative criteria for their definition
(e.g., Dariush et al. \cite{dar07}) and the concept of fossil
clusters for massive systems (e.g., Cypriano et al.  \cite{cyp06})
have been proposed.  Moreover, studies based on $N$-body numerical
simulations have suggested that many systems go through an optical
fossil phase during their life (e.g., von Benda-Beckmann et
al. \cite{von08}; Cui et al.  \cite{cui11}).

Recent observational results are often in contrast with the previous
results that found no particularly high mass concentration (Democles
et al. \cite{dem10}) and no special X-ray properties (12 fossil
systems, Voevodkin et al. \cite{voe10}; 10, Proctor et
al. \cite{pro11}; 17, Harrison et al. \cite{har12}).  Instead, Proctor
et al. (\cite{pro11}) claim atypical richnesses and optical
luminosities, but this has not been found by Voevodkin et
al. (\cite{voe10}) and Harrison et al. (\cite{har12}).  Recent studies
of fossil systems have also challenged the former conclusions of an
early formation of their BCGs from a gas-rich merger. Analyzing the
photometric and structural properties of BCGs in fossil systems, La
Barbera et al.  (\cite{lab09}, 25 fossil systems) and M\'endez-Abreu
et al. (\cite{men12}, 20 fossil systems, hereafter Paper II) have
found that they are similar to bright field ellipticals and to normal
cluster BCGs, respectively.  Finally, there is sparse evidence of a
few fossil systems far from being dynamically relaxed (e.g., Harrison
et al. \cite{har12}; La Barbera et al. \cite{lab12}; Miller et
al. \cite{mil12}).

Summarizing, there is still an open discussion on the real nature and
origin of fossil systems. For instance, on the basis of their
observational results, Harrison et al. (\cite{har12}) suggest that
fossil systems formed rather early and their galaxies represent the
end products of galaxy mergers, while Proctor et al. (\cite{pro11})
question the merging scenario, suggesting that the cannibalism of
bright central galaxies is not a convincing explanation for the
magnitude gap. Possible causes of the discrepancies among
observational results reported in the literature might be connected
with the use of very small samples, the presence of possible biases in
the estimates of physical quantities, or inhomogeneities in the
treatment of data of fossil and normal systems.

In 2008 we started a large observational program of fossil systems,
the FOssil Group Origins (FOGO) project (Aguerri et
al. \cite{agu11}; hereafter Paper I). The aim of this project is to
carry out a systematic, multiwavelength study of a sample of 34 fossil
group candidates identified by Santos et al. (\cite{san07}, hereafter
S07); here each system is denoted by FGS01, FGS02, etc., according to
the S07 list. The FOGO project was awarded time as International Time
Programme (ITP08-4 and ITP09-1) at the Roque de los Muchachos
Observatory for a total of 52 nights of observations.  Most optical
and NIR observations were performed during the period November
2008--May 2010 at the TNG, NOT, WHT, and INT telescopes. The
spectroscopic observations went on until April 2012 thanks to
additional time awarded at TNG through the Spanish and Italian Time
Allocation Committees. The catalog is described in the companion study
by Zarattini et al. (\cite{zar13}; hereafter Paper IV).

The first group we analyzed, RX J105453.3+552102 (FGS10 in the
S07 catalog), is a special system, because it is already a very
massive, relaxed galaxy cluster ($M\sim 1$ \mquii) at
$z=0.47$. Contrary to the findings of previous works that claim a boost in
the X-ray properties in fossil systems, FGS10 is quite normal
as shown by its position in the $L_{\rm opt}$--$L_{\rm X}$ plane (see
Paper I). Here we present our statistical results for 28 out of the 34
groups catalogued as fossils by S07.  We have taken care to apply
homogeneous procedures to the fossil and comparison systems and, in
particular, we have computed consistent optical luminosities.  Our
present analysis is mainly based on optical data from the Sloan
Digital Sky Survey Data Release 7 (hereafter SDSS-DR7, Fukugita et
al. \cite{fuk96}; Gunn et al. \cite{gun98}; Abazajian et
al. \cite{aba09}) and X-ray data from the {\em ROSAT} All Sky Survey
(RASS, Voges et al. \cite{vog99}). We have also used the results of paper
IV and, in particular, our check of the fossil classification of
the S07 objects.

This paper is organized as follows. We describe the S07 sample and the
comparison sample in Sect.~2. We detail the computation of X-ray and
optical luminosities in Sects.~3 and 4.  We devote Sect.~5 to the
comparison between fossil and normal systems in the $L_{\rm
  opt}$--$L_{\rm X}$ plane.  We discuss our results and present our
conclusions in Sect.~6.
 
Unless otherwise stated, we indicate errors at the 68\% confidence
level (hereafter c.l.).  Throughout this paper, we use $H_0=70$ km
s$^{-1}$ Mpc$^{-1}$ and $h_{70}=H_0/(70$ km s$^{-1}$ Mpc$^{-1}$) in a
flat cosmology with $\Omega_{\rm m} =0.3$ and $\Omega_{\Lambda}=0.7$.
Unless otherwise stated, all cosmology-dependent quantities that we
take from the literature are rescaled to our adopted cosmology.

\section{Samples of fossil and normal galaxy systems}
\label{cat}

Santos et al. (\cite{san07}) list 34 galaxy systems in the range of
redshifts $0.03<z<0.49$ catalogued as fossil group candidates.  These
systems were obtained as the result of a cross-match of the positions
of all luminous galaxies with measured spectroscopic $z$ in the
SDSS-Early Data Release (LRG catalogued by Eisenstein et
al. \cite{eis01}) with sources in the RASS with extended emission and
having a galaxy/{\em ROSAT} source distance of less than
$0.5^{\prime}$.  Only LRGs with magnitude $r$ $<19$ and
elliptical-type were considered by S07. In addition, S07 looked for
the LRG companions in the SDSS-DR5, taking objects classified as
galaxy within a radius of 0.5 \hh, and having the spectroscopic
redshift $z_{\rm spec}$, if available, $|z_{\rm spec}-z_{\rm LRG}| <
\Delta z=0.002$ or the photometric redshift $z_{\rm phot}$, $|z_{\rm
  phot}-z_{\rm LRG}|< \Delta z=0.1$. The systems so constructed were
included in the S07 catalog if the magnitude difference between the
LRG (i.e., the BCG of the system) and the second-brightest member was
$\Delta m_{12}\ge 2$ mag.  The authenticity of their fossil
classification is widely analyzed and discussed in Paper IV, where we
used new deep $r$-band images and optical spectroscopy information.
Out of 34 S07 objects, 15 showed to be genuine fossil groups having
$\Delta m_{12}\ge 2$ mag or $\Delta m_{14}\ge 2.5$ mag within
$0.5R_{200}$. The other 19 objects are either not fossil or their
fossil nature cannot be assessed with available data.  In the present
study, all the S07 objects are considered, except FGS19 because it was not
entirely sampled by the SDSS-DR7, and FGS11, FGS15, FGS28, FGS29, and
FGS32 because a significant peak was not detected by our analysis of
the 2D galaxy distribution (see Sect.~\ref{2d}). Our ALL-FGS
sample includes the remaining 28 S07 systems, 12 being confirmed
fossil systems (hereafter the CONF-FGS sample; the complementary
sample of 16 objects is denoted by NOCONF-FGS). The
NOCONF-FGS sample is used as the comparison sample.

As a more extended comparison sample, we considered a sample of
normal galaxy systems, i.e., galaxy systems not specifically
selected on the basis of their $\Delta m_{12}$
values. Specifically, we considered a subsample of 102 systems in the
redshift range $0<z<0.5$ extracted from the RASS-SDSS galaxy cluster
survey (Popesso et al. \cite{pop04}, hereafter P04). Following the P04
list, here each system is denoted by CL01, CL02, etc. The RASS-SDSS
survey lists 114 galaxy systems in the range of redshifts
$0.003<z<0.78$ and covers a wide range of masses from groups of
$10^{12.5}$ $h_{70}^{-1}$ $M_{\odot}$ to massive clusters of $10^{15}$
$h_{70}^{-1}$ $M_{\odot}$.  It comprises all the X-ray selected
objects already observed by the SDSS up to February 2003. The reason
for using this sample for the comparison is threefold: it is quite
large; it is based on the RASS and SDSS surveys, the same data
sources used by S07; and it has been used by P04 to analyze optical
luminosities, and thus several technical points have already been
outlined and properly verified by P04 and following studies. From the
114 RASS-SDSS clusters we do not consider: the four systems classified
as FGS by S07 (CL005=FGS02=Abell 267, CL017=FGS05=Abell 697;
CL103=FGS30=ZwCl 1717.9+5636;CL105=FGS31), the five systems with X-ray
luminosity listed as 0.00 by P04 (CL018; CL050; CL052; CL055; CL070 of
which the last four have redshift $z<0.01$), the other two systems
with $z<0.01$ (CL082; CL083), and the system with the highest redshift
(CL044 at $z=0.784$).  We obtained a sample of 102 systems (hereafter
the CL sample) with $0.01<z<0.46$, i.e., in the same redshift range of
S07 FGSs.  The X-ray luminosity distributions of ALL-FGSs and CLs are
not statistically different (see the Sect.~\ref{estlx} for
$L_{\rm X}$ computation).  Since the X-ray luminosity is a proxy for
the mass of galaxy systems, the FGS sample is expected to span a 
  range of masses that is comparable to that of the CL sample.
However, their $z$ distributions differ at the $>99\%$ level according
to the Kolmogorov--Smirnov test (hereafter 1DKS-test; see, e.g.,
Ledermann \cite{led82}). The FGS $z$ distribution is picked at
higher values ($\Delta z \sim 0.1$).  Thus, we expect that FGSs are
somehow less optically contrasted onto the sky than CLs.

The FGS and CL samples are listed in Tables \ref{tabfg} and
\ref{tabpop1}. For each FGS, Table~\ref{tabfg} lists notes about
  their classification (Col.~2); the center (R.A. and Dec.) and
redshift $z$, as taken from S07 and referring to the BCG (Cols.~3 and
4); the X-ray luminosity, $L_{\rm X}$, in the (0.1-2.4) keV band
(Col.~5); the radius $R_{500}$, and the optical $r$-band luminosity
computed within $R_{500}$, $L_{\rm opt}(<R_{500})$ (Cols.~6 and 7);
$L_{\rm opt}(<0.5R_{200})$ being $R_{200}=1.516\times R_{500}$
(Col.~8); and additional information (Col.~9).  The listed values of
$L_{\rm X}$, $R_{500}$, $L_{\rm opt}$ are derived in the following
Sections.  For each CL, Table~\ref{tabpop1} lists the same properties
where the CL centers and redshifts are taken from P04, as well as
the X-ray luminosity values (here converted to our adopted cosmology).

\input{tabfg}

\input{tabpop1}

\section{X-ray luminosities}
\label{lx}

\subsection{X-ray luminosity estimates}
\label{estlx}

Our reference values for the X-ray luminosities of the CL sample are
those computed by P04 and listed in their Table~1.\footnote{In P04,
  $L_{\rm X}$ values are listed for H$_0=50$ km s$^{-1}$Mpc$^{-1}$,
  $\Omega_{\rm m}=1$, and $\Omega_{\Lambda}=0$.}  Santos et
al. (\cite{san07}) list the L$_{\rm X}$ values of FGSs in the (0.5--2)
keV band as computed from {\em ROSAT} count rates. A quick comparison
between the values of a few FGSs, which are also well-known clusters
(e.g., Abell 267 = FGS02 and Abell 697 = FGS05), with the published
values (e.g., B\"ohringer et al. \cite{boe00}), shows that they are
underestimated by a factor of $\sim$2.  We thus decided to recompute
X-ray luminosities for the full S07 sample.

For each FGS, we considered the counts from the RASS Bright Source
Catalog (RASS-BSC; Voges et al. \cite{vog99}) or, alternatively, from the RASS
Faint Source Catalog (RASS-FSC; Voges et al. \cite{vog00}), which are in the
broad band 0.1--2.4 keV. We used the total Galactic column density
($N_{\rm H}$) as taken from NASA's HEASARC $N_{\rm H}$
tool\footnote{http://heasarc.gsfc.nasa.gov/cgi-bin/Tools/w3nh/w3nh.pl}
and the redshift $z$ as listed by S07.

The computation of the flux was made by using an iterative procedure
based on the PIMMS\footnote{At ftp:${\rm
    //legacy.gsfc.nasa.gov/software/tools/pimms4\_3.tar.gz}$.}
software available at NASA's HEASARC tools (Mukai \cite{muk93}).
We adopted the plasma model, a metal abundance of 0.4, and, at the
first step, a starting value for the temperature $kT_{\rm X}=2$
keV. The resulting unabsorbed flux is slightly corrected to take into
account the flux coming from the outer regions ($\times 1.08$, which
is the mean value in the NORAS clusters, B\"ohringer et
al. \cite{boe00}). This flux was used to compute a first estimate of
the X-ray luminosity (in the 0.1-2.4 keV band).  We used the X-ray
luminosity to compute an estimate of the temperature through Eq.~4 in
B\"ohringer et al.  (\cite{boe00}) derived from the
luminosity-temperature relation in Markevitch (\cite{mar98}) and used
for the NORAS clusters $kT_{\rm X}=2.34\ {\rm keV}\, L_{\rm
  X,44,H0=50}^{1/2}$, where $L_{\rm X,44,H0=50}$ is the X-ray
luminosity in units of $10^{44}$ erg s$^{-1}$, in the 0.1-2.4 keV
band, and in the B\"ohringer et al.  (\cite{boe00}) cosmology. This
temperature and the redshift of the system were used to compute the
K-correction (B\"ohringer et al. \cite{boe04}; see their Table~3).
The K-corrected X-ray luminosity allowed us to obtain a new
estimate of the temperature, which could be used as the new
starting value in the PIMMS procedure. The second iteration of the
procedure is enough to converge to the final luminosities and
temperatures, $L_{\rm X}$ and $T_{\rm X}$. Throughout the paper, these
$L_{\rm X}$ estimates are our reference values for the FGS sample and
are listed in Table~\ref{tabfg}.  The question of the level of
homogeneity of these estimates with those taken from P04 for the CL
sample is addressed in Sect.~\ref{errlx}.

\subsection{Characteristic radius estimates}
\label{rad}

In Sect.~\ref{lo} we present our estimation of reference optical
luminosities as computed within a radius of $R_{500}$. We also estimated
luminosities within $0.5R_{200}$ for useful comparison with other
authors.  The use of a characteristic radius is suggested in order
to treat comparable regions for galaxy systems of different masses.  For
each system, we computed $R_{500}$ using Eq.~2 in B\"ohringer et
al. (\cite{boe07}),

\begin{equation}
R_{500}=0.753\ {\rm Mpc} \ h_{100}^{-0.544} E(z)^{-1} L_{\rm X,44}^{0.228},
\end{equation}

\noindent where $E(z)=h(z)/h_0$ and $L_{\rm X,44}$ is the X-ray
luminosity in units of $h_{70}^{-2}10^{44}$ erg s$^{-1}$ (in the
0.1--2.4 keV band). This equation is based on the $R_{500}$--$T_{\rm
  X}$ relation by Arnaud et al. (\cite{arn05}; see details in the
original papers). Following Arnaud et al. (\cite{arn05}; see their
Table~2 for whole cluster sample results), we computed $R_{200}=1.516
\times R_{500}$, in agreement with numerical simulations where
$R_{200}/R_{500}\sim 1.5$ for the typical halo concentration parameter
$c=5$ (Yang et al. \cite{yan09}).  The median value of $R_{500}$ for
FGSs (and CLs) is $\sim 0.9$ \hh.

\subsection{$L_{\rm X}$ estimates: uncertainties and homogeneities}
\label{errlx}

We adopted the  value $\epsilon_{L{\rm
    x}} = 20\%$ for CLs, taken from P04 as a typical $L_{\rm X}$
uncertainty. In the case of FGs, we used the count error listed by
RASS-BSC/FSC and computed the relative error.  The same relative error
was assumed for $L_{\rm X}$ ($\epsilon_{L{\rm x}}\sim 25\%$; median
value).

The X-ray luminosities computed by P04 were not obtained using the
RASS-BSC/FSC counts, but rather with the counts estimated through the
GCA method (B\"ohringer et al. \cite{boe00}, NORAS clusters).
B\"ohringer et al. (\cite{boe00}) have pointed out that RASS-BSC/FSC
underestimate counts, probably because of the design of the source analysis
technique used for RASS (see their Fig.~11b).  To check the effect of
this on our $L_{\rm X}$ estimates, we recomputed X-ray
luminosities for 100 out of 102 CLs following the same procedure we
used for FGSs (see Sect.~\ref{estlx}), these alternative estimates being
labelled as $L_{\rm X,BSC/FSC}$. For two of the 102 CLs in our
comparison sample we failed to find any RASS-BSC/FSC source within
5\arcm from the P04 center and we did not consider them. We found that
the difference of the two alternative estimates strongly depends on
whether the system is recognized as an extended source or not by the
RASS-BSC/FSC catalogs; the extended emission is one of the selection
criteria required by S07.  Among the 100 CLs, 67 and 33 systems are classified
as extended and nonextended sources, respectively.  For the 33
nonextended sources, we confirm a large systematic difference, finding
$L_{\rm X,P04}/L_{\rm X,BSC/FSC}=2.6$ (median value). For the 67
extended sources, the two alternative estimates are only slightly
different, $L_{\rm X,P04}/L_{\rm X,BSC/FSC}=1.21$. The presence of a
systematic (although small) difference led us to also consider two
alternative approaches when comparing FGSs and CLs (see
Sect.~\ref{cfr}): i) using $L_{\rm X,BSC/FSC}$ for CLs (only the 67
extended systems are considered), and ii) applying a correction to the
FGS X-ray luminosities determined in Sect.~\ref{estlx} in such a way as
to more closely resemble those listed by P04 for CLs. The correction was
obtained by fitting $L_{\rm X,P04}$ vs. $L_{\rm X,BSC/FSC}$ for the
sample of the 67 extended CLs.  The direct regression line,
recommended to predict the value of the $y$ variable (see, e.g., Isobe
et al. \cite{iso90}), is ${\rm log}(L_{\rm
  X,44,P04})=0.136+0.865\cdot{\rm log}(L_{\rm X,44,BSC/FSC})$, where
$L_{\rm X,44}$ is the X-ray luminosity in units of
$h_{70}^{-2}10^{44}{\rm erg\,s}^{-1}$.  The corrected luminosities
for FGSs, hereafter $L_{\rm X,corr}$, are obtained from the values
computed in Sect.~\ref{estlx} using the right-hand side of the above
equation.

\section{Optical luminosity estimates}
\label{lo}

\subsection{Galaxy catalogs}
\label{sdss}

The galaxy catalogs were obtained from the SDSS-DR7. For each galaxy
system, we considered objects within a circular region with a radius
of 30\arcm positioned on the center listed by S07 (P04 for CLs). Only
objects classified as extended and not containing one or more
saturated pixels were selected.  The last constraint is required to
reject stars classified as bright galaxies (e.g., Yasuda et al.
\cite{yas01}).  We always considered only objects labeled ``PRIMARY''
(see Yasuda et al.  \cite{yas01} for more details).  As a further
check, we have also looked at objects classified as extended and
saturated objects, but that are real galaxies having $cz>1000$
\kss. The inclusion of these (few) objects---almost always nonmember,
foreground galaxies---would change the average observed luminosity for
only 8 of the 136 analyzed systems. For the sake of completeness of
our catalogs, we decided to include only three galaxies: the BCGs of
FGS21, CL013, and CL100; other differences are negligible.

In order to compare with previous works in the literature, we
considered SDSS $r$-band magnitudes.  The SDSS photometry of
point-like sources is nominally $95\%$ complete down to a model
magnitude $r=22$ (Stoughton et al. \cite{sto02}) and the
star/galaxy classification is still reliable down to $r\sim
21.5$ (Lupton et al. \cite{lup01}; see also Capozzi et
al. \cite{cap09}).  Accordingly, we adopt here a limiting
magnitude of $r\sim 21.5$ for the entire SDSS catalog.

We used dered magnitudes (hereafter $m_{r}$),
i.e., model magnitudes already corrected for the Galactic absorption
(hereafter ${\cal A}_r$). We applied both K-dimming and evolutionary
correction. We used the K-correction ${\cal K}_r(z)$, supplied by
Fukugita et al. (\cite{fuk95}), for elliptical galaxies, assuming that
the main population of galaxy systems in our samples are the old
elliptical galaxies at the system redshift (see also P04). We also used
the evolutionary correction ${\cal E}_r(z)=0.86z$ from Roche
et al. (\cite{roc09}), which is typical for elliptical galaxies. The
absolute magnitude is defined as

\begin{equation}
M_r=m_{r}-25-5{\rm log}_{10}(D_L/1{{\rm Mpc}})-{\cal K}_r(z)+{\cal E}_r(z),
\end{equation}

\noindent where $D_L$ is the luminosity distance in \hh.  

\subsection{Checking the 2D galaxy distribution of S07 objects}
\label{2d}

 While RASS-SDSS clusters are well-studied systems in the literature
 in both the X-ray and the optical wavelengths, this is not true for
 all S07 objects.  Using NED\footnote{The NASA/IPAC Extragalactic
   Database (NED) is operated by the Jet Propulsion Laboratory,
   California Institute of Technology, under contract with the
   National Aeronautics and Space Administration.}  we have found that
 23 S07 objects have been clearly identified as galaxy systems in one
 or more optical cluster/group catalog(s) based on photographic plates
 or SDSS and few of them are well-known systems. On the contrary, 11
 S07 objects have no such identification, the closest galaxy system
 being more distant than 4\arcmin. For each FGS object, we analyzed
 the galaxy distribution in the region around the BCG through the 2D
 DEDICA method, which is an adaptive-kernel method (Pisani
 \cite{pis93} and \cite{pis96}; see also, e.g., Girardi et
 al. \cite{gir11} for a recent application).  This method of density
 reconstruction gives as output the list of density peaks, their
 significance, density, and richness, as well as the relative
 membership.  To minimize the effect of foreground/background galaxies we
 only worked on galaxies (hereafter likely members) having a color
 close to that of BCG, i.e., $|(r-i)-(r-i)_{\rm BCG}|\leq 0.2$ (see
 also Harrison et al. \cite{har12}), and having magnitude $M_r<-19$,
 in order to sample the luminosity function down to $\sim
 M_r^*+3$ mag, if possible, but not considering fainter galaxies.

\begin{figure*}
%\centering
\includegraphics[width=9cm]{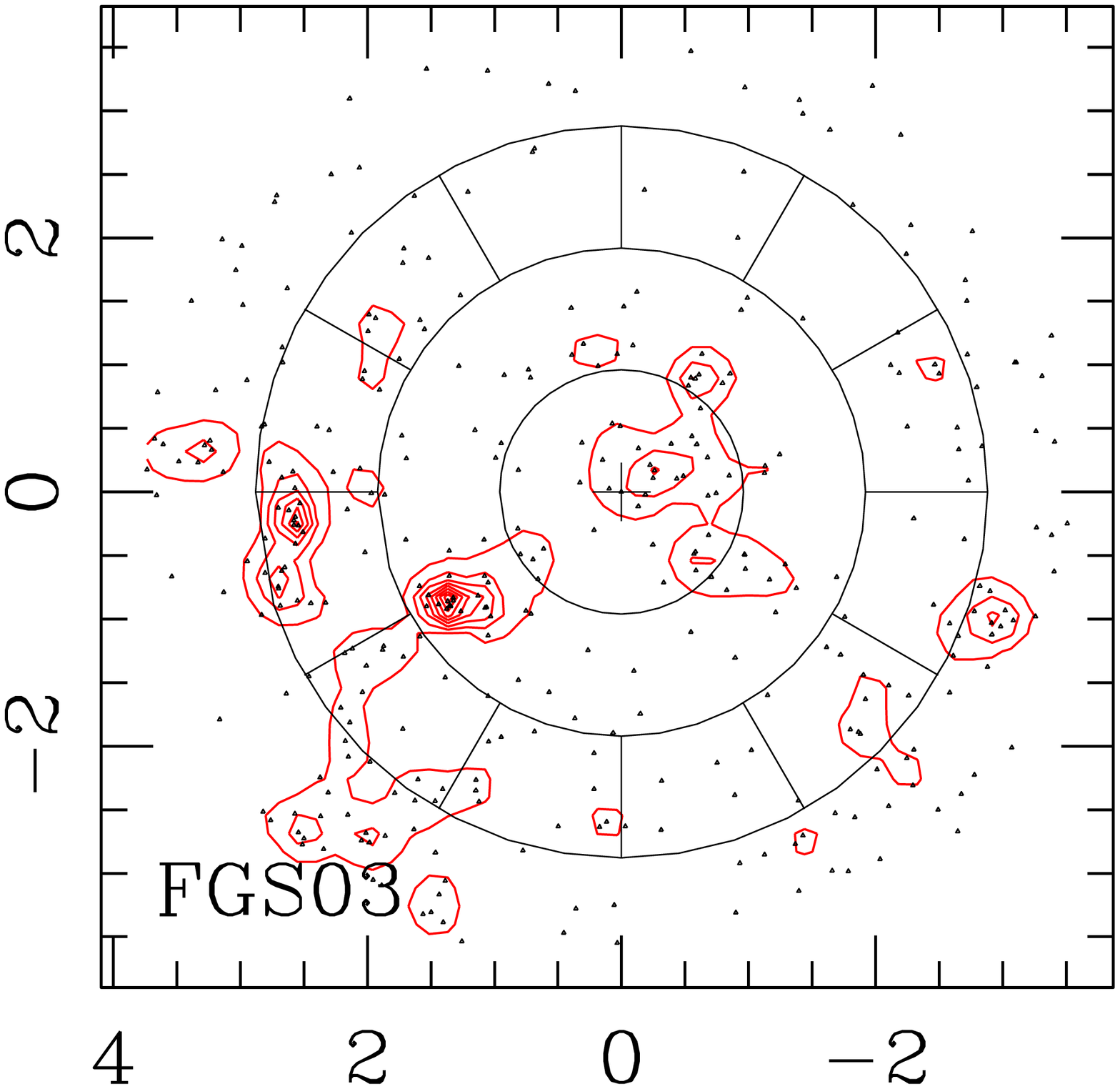}
\includegraphics[width=9cm]{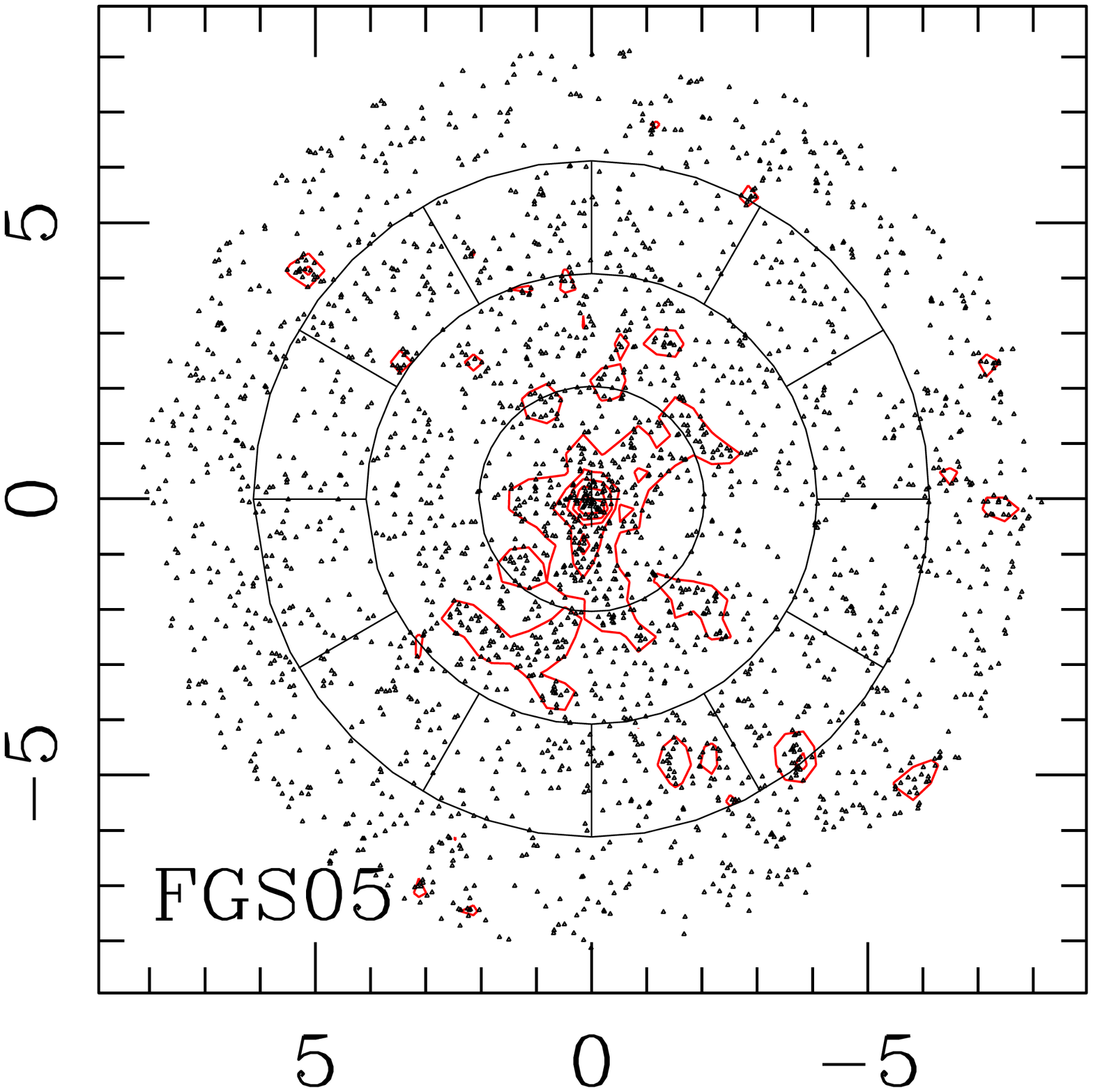}
\includegraphics[width=9cm]{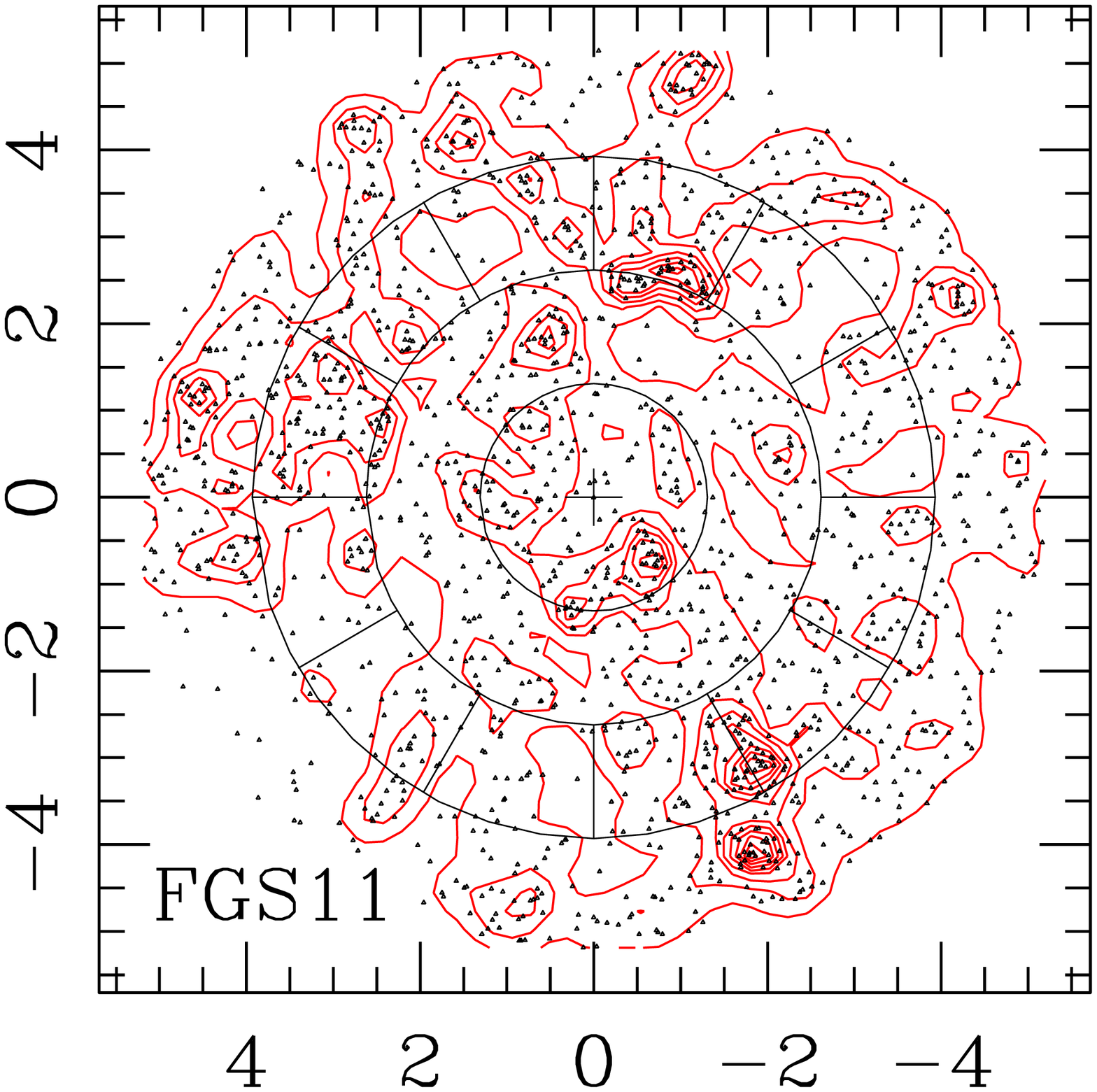}
\includegraphics[width=9cm]{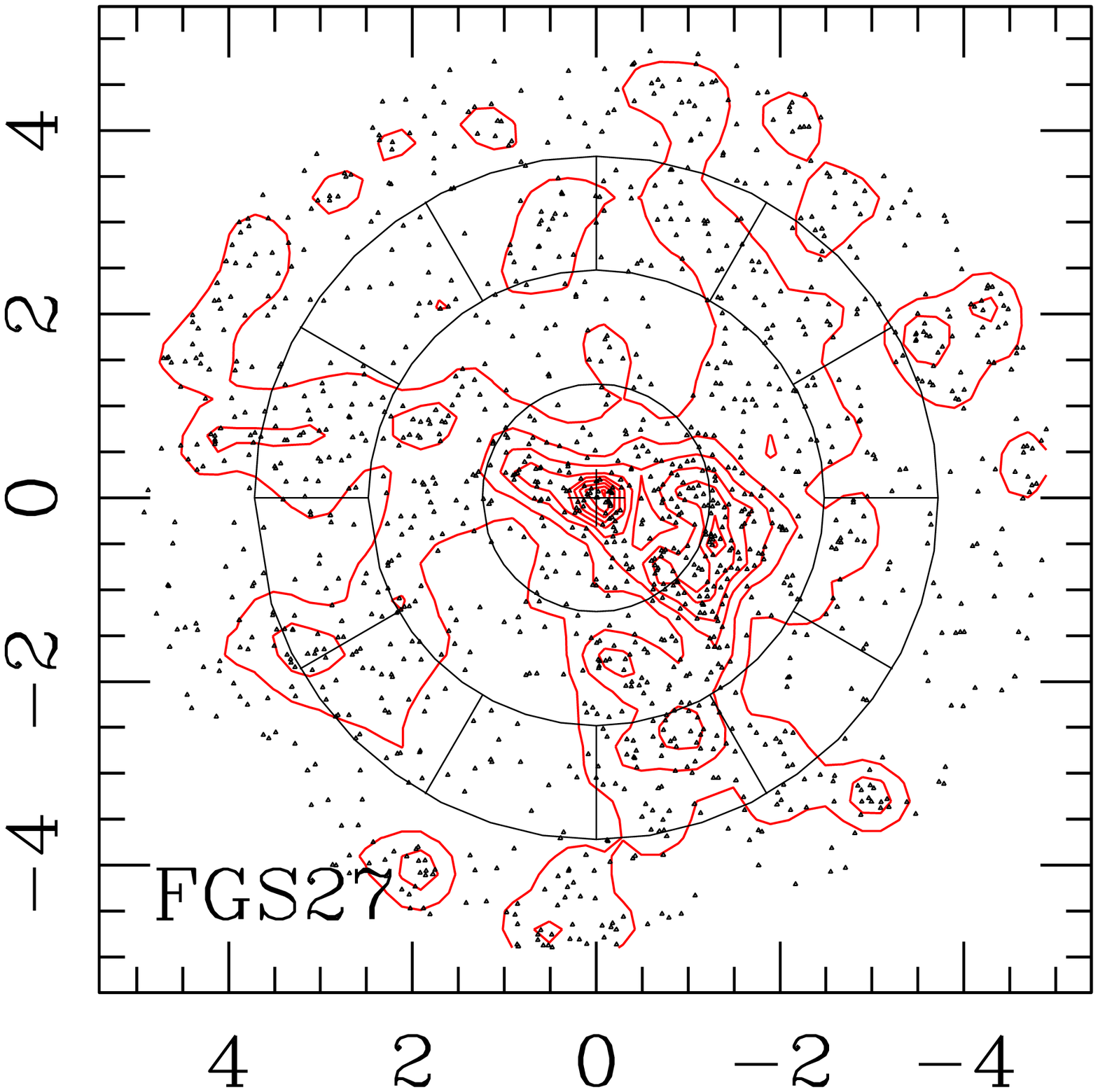}
\caption
{Projected spatial distribution and 2D-DEDICA isodensity contours of
  SDSS galaxies with $|(r-i)-(r-i)_{\rm BCG}|\leq 0.2$ and $M_r<-19$
  in a few FGSs spanning a wide range of appearances (see text). The plots are
  centered on the S07 BCGs (marked with black crosses). The inner circle
encloses the region within $R_{200}$. The outer two circles
  enclose the regions within 2$R_{200}$ and 3$R_{200}$.  The sectors  
used for the computation of the local field are displayed.  Units on
  the axes are in \hh.  }
\label{fig2d}
\end{figure*}

For most S07 objects, there is an excellent match between the
location of the BCG and the densest density peak in the whole 
$2R_{200}$ region [hereafter Ipeak($2R_{200}$)].  In a sample of 24 S07
systems, the median distance between the BCG and the density peak location is
$d\sim 80$ \kpc $\sim 0.05 R_{200}$.  In the above cases the presence
and identification of a galaxy system is outstanding and, with the
exception of FGS13, they all already have a corresponding system
in one or more published optical cluster catalogs.

In other cases (FGS03, FGS08, FGS10, FGS32, FGS34), the BCG does not
correspond to the Ipeak($2R_{200}$), but to the densest peak
within $R_{200}$, Ipeak($R_{200}$), often separated by a great
distance, e.g., $d\lesssim 0.5 R_{200}$ in FGS03 and FGS10. This means
that these FGSs can be strongly contaminated by a very dense galaxy
system that is close enough. However, FGS03, FGS08, FGS10, and FGS34
are very rich [Ipeak($R_{200}$) is richer or comparable to
  Ipeak($2R_{200}$)], and/or have $z$-data to support the existence of
an extended galaxy system (see Paper IV), and appear sufficiently
contrasted with respect to the field (see our local field computation
in Sect.~\ref{estlo}); the noticeable negative exception is thus
FGS32.  In other cases, no significant peak can be detected within
$R_{200}$ (FGS28), or Ipeak($R_{200}$) is far from the BCG, i.e.,
$d>0.5 R_{200}$ (FGS11, FGS15, FGS18, FGS29). However, in the case
of FGS18, the BCG is closely located to a significant secondary peak,
IIpeak($R_{200}$), somewhat contrasted with respect to the field
around it.  Summarizing, we did not consider FGS11, FGS15, FGS28,
FGS29, FGS32.  All the rejected objects have no corresponding system
in published optical cluster catalogs.

The objects listed in S07 span a wide range of morphological
appearances; some are very dense, concentrated, and isolated systems,
while others are very substructured and/or surrounded by a rich large
scale structure.  Figure~\ref{fig2d} shows a few examples of the
2D-DEDICA contour maps: FGS05=Abell 697, probably the most massive
system in the S07 sample, is well isolated in the 2D space (but not
confirmed to be a fossil system); FGS27, a massive fossil cluster;
FGS03, a poor nearby fossil group, but just acceptable enough to be
part of our analysis (see above); and FGS11, a S07 object not
considered in this study.  With the present data, we cannot be
definitive about the nature of the rejected objects.  We suspect that
they might not correspond to an extended system (or that they are only
poor subsystems).  The S07 identification of extended systems based
only on the RASS-BSC and FSC definition of extended sources might not
always be reliable.  For instance, out of six fossil groups identified
by La Barbera et al. (\cite{lab09}) in a similar way, the following
XMM X-ray data analysis shows that one does not have an extended
emission, and another is at the border of a real extended system (La
Barbera et al.  \cite{lab12}).  Alternatively, the rejected objects
might simply be too poorly contrasted in the sky.  In either case, we
were not able to perform a reliable computation of the optical
luminosity.

In summary, considering the rejection of the five FGSs with no clear
identification in the galaxy distribution and FGS19 (which is not fully sampled
by SDSS-DR7), our working sample is formed of a sample of 28 FGSs (the
ALL-FGS sample), 12 of which are confirmed fossil systems (the
CONF-FGS sample), while the complementary sample of 16 objects is
called NOCONF-FGS.

\subsection{Computing $L_{\rm opt}$}
\label{estlo}

We computed $L_{\rm opt}$ within $R_{500}$ (and $0.5R_{200}$)
following standard procedures for photometric samples (e.g., Girardi
et al.  \cite{gir00}; P04). In particular, P04 suggests that the
count-based $L_{\rm opt}$ estimation has to be preferred to the
fit-based one in the study of the correlation between optical and
X-ray properties (see their Sect.~5.3); our procedure is of the
count-based type.

Observed cluster/group luminosities, $L_{\rm obs}$, were obtained by summing the
individual absolute luminosities of all galaxies and assuming the
absolute magnitude in the $r$-band for the Sun as
$M_{\odot,r}=4.68$ (as listed by SDSS).

The observed luminosity needs to be corrected for foreground/background
contamination, which is the largest source of uncertainty in these kinds
of estimates (see, e.g.,  P04).  Two approaches can be used for the
statistical subtraction of the galaxy background: the local and the
global backgrounds.  The limitation of the global background is
that local fluctuations of the luminosity field are not taken into
account. The alternative is the local background method, which 
is limited by the Poisson uncertainty of the counts.  As
for FGSs, we decided to compute an individual local field.  For each
FGS we extracted from SDSS-DR7 the catalog of galaxies in the annulus
between $2R_{200}$ and $3R_{200}$ in such a way that the galaxy background
has been estimated outside the system, but still locally. However, one
risks obtaining a local field contaminated by close companion galaxy
systems.  To overcome this problem, the annulus was divided into 12
sectors, each sector having an area similar to that within $R_{500}$;
the sectors containing 2D-DEDICA contour levels indicating a
relative density $> 30\%$ with respect to the FGS peak were not
considered (as above, the DEDICA analysis was applied to the likely
members, see Sect. \ref{2d}).  We also did not take into account those
sectors not fully sampled by SDSS data. The surviving ${\cal N}$
sectors were used to compute the local counts for each FGS.

\begin{figure}
\centering
\resizebox{\hsize}{!}{\includegraphics{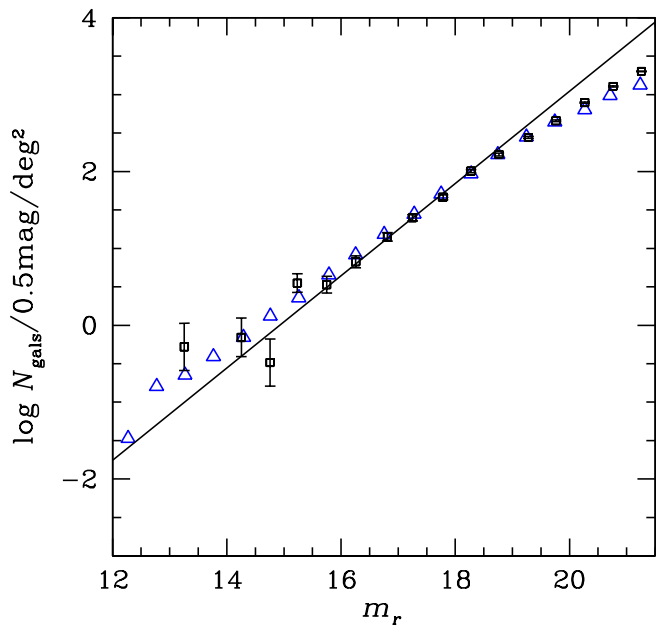}}
\caption
{Average field counts (squares) compared to those by P04 (blue
  triangles). The error bars represent 1sigma Poisson errors. The solid line is
  the fit for the Euclidean geometry.}
\label{figcounts}
\end{figure}

The local counts in the magnitude bins for each FGS were then averaged
all together.  These average counts, estimated using a global area of
$\sim 5$ deg$^2$, agree rather well with those of P04 (see
Fig.~\ref{figcounts}).  The line in Fig.~\ref{figcounts} shows a fit
to the galaxy counts-magnitude relation expected in a homogeneous
universe assuming Euclidean geometry for a 3D space
${\rm log}[N (m_r)] = {\rm log}(A)+0.6(m_r-16)$ (see Yasuda et
al. \cite{yas01}). We obtained $A=4.41$ (0.5 mag)$^{-1}$ deg$^{-2}$
using four points in the range $16<m_r<18$ (at $m_r\sim 15.75$ the
number of galaxies is already quite small, $N=16$).  We define our
global background as the combination of our average counts and
the Euclidean fit for $m_r>15.75$ and $m_r<15.75$, respectively.  A
posteriori, we verified that, on average, the different corrections do
not significantly affect the cluster/group luminosity estimation: we
found that $L_{\rm opt,loc-back}/L_{\rm opt,glob-back} = 1.01 \pm
0.03$ with rms$=0.18$.

For each system, we computed the corrected luminosity, $L_{\rm corr}$,
by subtracting the background luminosity $L_{\rm back}$ obtained from
the background counts rescaled to the area of the system $L_{\rm
  corr}=L_{\rm obs}-L_{\rm back}$.  Before the field subtraction, the
field galaxies of each system were treated in the same way as the
respective member galaxies, i.e.,  we applied the same conversion in
absolute magnitudes of Eq.~2.  Corrected counts $N_{\rm corr}$ are
then obtained in a similar way.  As for the FGS sample, the typical
correction is $\sim 35\%$ (median value) with the worst case having a
$75\%$ correction (FGS12).  

In the case of the CL sample, we applied the above procedure
adopting the global background; P04 has shown that the
luminosity difference using a local or global background is smaller
than the statistical error. The typical correction ($\sim 25\%$) is
smaller than in the case of FGSs, in agreement with the fact that FGSs
are expected to be somewhat less contrasted on the sky than the CLs.

In order to obtain the total optical cluster/group luminosity $L_{\rm opt}$, we need
to add the contribution of the galaxies below the magnitude completeness
limit. To compute this contribution we adopted the usual Schechter
(\cite{sch76}) form for the cluster luminosity function (LF)
obtaining

\begin{equation} 
L_{\rm opt}=L_{\rm corr}+\Phi^* L^*\int^{L_{\rm lim}/L^*}_{L_{\rm min}/L^*}
x^{1+\alpha} e^{-x}dx,
\end{equation} 

\noindent 
where $L_{\rm lim}$ is the luminosity corresponding to the limiting
magnitude; $L_{\rm min}$ corresponds to a cutoff for the minimum
galaxy luminosity (here we adopt $L_{\rm min}=10^{-4}L^*$); and $L^*$,
$\alpha$, and $\Phi^*$ are the parameters of the LF.

We adopted the LF parameters determined by Popesso et
al. (\cite{pop05}), i.e., the $L^*$ value corresponding to the
absolute magnitude $M_{r}^*=-22.12+5$log$h_{70}$ and $\alpha=-1.30$ as
listed in the first part of their Table~2 (second line).  Following
previous studies (Lumsden et al. \cite{lum97}; Girardi et al.
\cite{gir00}), the $\Phi^*$ parameter is determined from the
(corrected) galaxy number counts in a magnitude interval around 
$M^*$ to obtain a more robust value. We used $N_{\rm corr}(-19,-23)$
computed for $-23 \le M_{r} \le -19$ to obtain

\begin{equation}
\Phi^*=N_{\rm corr}(-19,-23)/\int_{L(-19)/L^*}^{L(-23)/L^*}x^{\alpha}e^{-x}dx,
\end{equation} 

\noindent where $L(-19)$ and $L(-23)$ are the luminosities
corresponding to absolute magnitudes of $M_{r}=-19$ and $M_{r}=-23$,
respectively.  If the absolute limiting magnitude is brighter than
$M_{r}=-19$, we take $L_{\rm lim}$ for the lower integration limit in
Eq.~4.  Owing to the extrapolation to faint magnitudes, the luminosity
increases by $\sim 10\%$ and $5\%$ (median values for FGSs and
CLs).  The obtained luminosity $L_{\rm opt}$ is considered our
reference optical luminosity.

\subsection{Uncertainties in $L_{\rm opt}$ estimates}
\label{errlo}

The foreground/background correction is the largest correction applied to
the observed luminosity and is the largest source of random error in
luminosity estimates. The comparison between $L_{\rm opt,loc-back}$
and $L_{\rm opt,glob-back}$ for FGSs suggests a $20\%$ estimate of the
luminosity uncertainties. We also estimated uncertainties for each
individual FGS using the field sectors adopted to compute the local
background.  For each FGS, we computed ${\cal N}$ optical luminosities
using the backgrounds as derived for the available ${\cal N}$ sectors:
the rms of their distribution (or half of the distribution range in
the case of ${\cal N}\le 4)$ was taken as an estimate of the
luminosity uncertainty for each individual FGS (on average $\sim
20\%$). Conservatively, for each FGS, the largest between this
individual estimate and the above global $20\%$ estimate is assumed to
be the statistical uncertainty due to the background (hereafter
$\epsilon_{\rm Lopt,back}$).  As for CLs, we assumed 
$\epsilon_{\rm Lopt,back} = 20\%$ , in agreement with
that directly estimated by P04.

The above uncertainty has been obtained for a fixed aperture. In
addition, for both FGSs and CLs, we had to take into account how the
uncertainty in the estimate of $R_{500}$ propagates to the $L_{\rm
  opt}$ computation.  First, the $R_{\rm 500}$ estimate is subjected to the
uncertainty in the $L_{\rm X}$ estimate: according to Eq.~1, the
formal error is small enough, i.e., $\epsilon_{r500} =1/0.228
\epsilon_{L{\rm X}} \sim 5\%$ and $8\%$ for CLs and FGSs. Second, one
should consider the intrinsic scatter (i.e., not due to measurement
errors) in the relations used to derive the value of $R_{\rm
  500}$ (Eq.~1 and those referred in the original papers) or, more 
generally, the intrinsic scatter between $R_{\rm 500}$ values estimated
from different observables.  This issue is connected to the cluster
mass calibration and its complete discussion is well outside the scope
of this paper. In this study we have considered the result of Zhang et
al. (\cite{zha11}), i.e., the presence of a $\sim 20 \%$ intrinsic
scatter in the relation between $R_{\rm 500}$, as determined from
X-ray observables, and velocity dispersion (see their Table~3).
Adding both sources of uncertainty, a $25\%$ error in the $R_{\rm
  500}$ estimate was considered.  The propagated uncertainty on
$L_{\rm opt}$, $\epsilon_{L{\rm opt,radius}}$, was computed as half
$|L_{\rm opt,r500+25\%}-L_{\rm opt,r500-25\%}|/L_{\rm opt}$, 
  where $L_{\rm opt,r500+25\%}$ and $L_{\rm opt,r500-25\%}$ are the
  luminosities in regions where the radius is $25\%$ larger
  and smaller than $R_{\rm 500}$. We obtained $\epsilon_{L{\rm
    opt,radius}}\sim 20\%$ (median value).  Summarizing, the estimate of
the total uncertainty on $L_{\rm opt}$ was then conservatively
computed as $\epsilon_{L{\rm opt}}= \epsilon_{L{\rm
    opt,radius}}+\epsilon_{L{\rm opt,back}} \sim 40\%$ (median value).

\section{Comparison between fossil and normal galaxy systems}
\label{cfr}

Here we present the comparison between CONF-FGSs and NOCONF-FGSs in the
$L_{\rm opt}$--$L_{\rm X}$ plane.  We also compared CONF-FGSs (and
ALL-FGSs) with CLs.  The first comparison has the advantage of being
only based on the S07 catalog and thus it handles a single selection
function, but it has the obvious drawback of being based on two small
samples. For the second comparison, we also explored the
possibility of using $L_{\rm X,BSC/FSC}$ for CLs or, alternatively,
$L_{\rm X,corr}$ for FGSs in such a way as to improve the homogeneity of
the comparison (hereafter homo- and corr-cases).  We note that X-ray
and optical luminosity estimates have always been consistently
determined, i.e., the radius used to compute $L_{\rm opt}$ is always
based on the corresponding X-ray luminosity estimate.  In practice, we
considered the following comparisons: 12 CONF-FGSs -- 16 NOCONF-FGSs;
12 CONF-FGSs -- 102 CLs; 12 CONF-FGSs -- 67 CLs (homo-case); 12
CONF-FGSs -- 102 CLs (corr-case); 28 ALL-FGSs -- 102 CLs; and 28 ALL-FGSs
-- 67 CLs (homo-case); 28 ALL-FGSs -- 102 CLs (corr-case).  We
considered optical luminosities within both $R_{500}$ and $0.5R_{200}$.

\begin{figure}
\centering
\resizebox{\hsize}{!}{\includegraphics{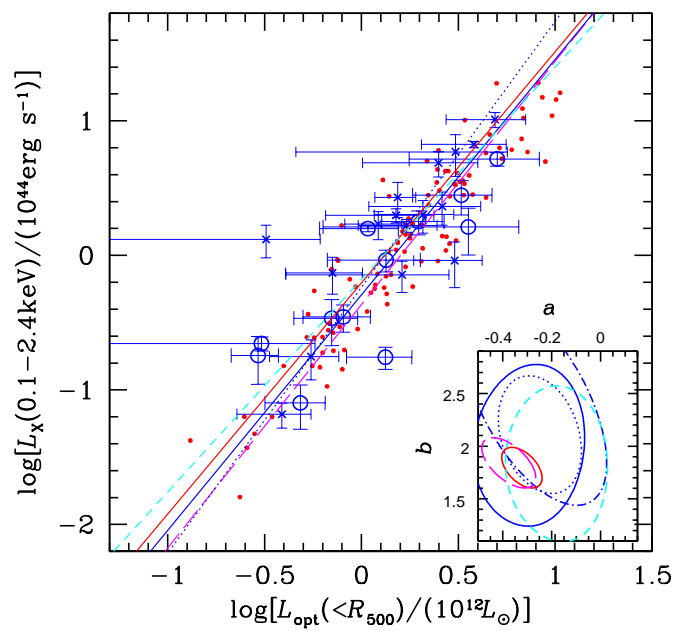}}
\caption
{X-ray luminosity vs. $r$-band optical luminosity for CONF-FGSs (blue
  circles), NOCONF-FGSs (blue crosses), and CLs (red dots). Error
    bars for the CL sample are omitted.  Reference values for $L_{\rm
    opt}$ and $L_{\rm X}$ are shown.  The blue solid and dotted lines
  indicate the fits for the CONF-FGS and ALL-FGS samples, and the red
  solid line is the fit for the CL sample.  The magenta long-dashed
  and cyan dashed lines are the fits when using alternative X-ray
  luminosity estimates: $L_{\rm X,BSC/FSC}$ values for the CL sample
  and $L_{\rm X,corr}$ values for the CONF-FGS sample. The inset plot
  shows the $90\%$ c.l. confidence ellipses corresponding to
  CONF-FGSs, NOCONF-FGSs, ALL-FGSs (solid, dot-dashed, and dotted blue
  curves), and CLs (solid red curve). Results for alternative X-ray
  luminosities are shown: $L_{\rm X,BSC/FSC}$ for CLs (magenta long
  dashed curve) and $L_{\rm X,corr}$ for CONF-FGSs (cyan dashed line).
}
\label{figlxlo}
\end{figure}

As a first approach we used the 2D Kolmogorov-Smirnov test (hereafter
2DKS-test, Fasano et al. \cite{fas87}, Press et al. \cite{pre92}),
which has the advantage of being a nonparametric test.  No significant
difference was detected. The comparison for our reference values
is shown in  Fig.~ \ref{figlxlo}.
We also performed the linear fit in the $L_{\rm opt}$--$L_{\rm X}$
logarithmic plane.  We used a maximum likelihood estimate of the
regression lines (see, e.g., Kendall \& Stuart \cite{ken79}; Press et
al. \cite{pre92}) to fit

\begin{equation}
{\rm log}(L_{\rm X,44})=a +b\cdot{\rm log}[L_{\rm opt,12}(<R_{500})],
\end{equation}

\noindent where $L_{\rm X,44}$ is the X-ray luminosity in units of
$h_{70}^{-2}10^{44}{\rm erg\,s}^{-1}$ and $L_{\rm opt,12}$ is the
optical luminosity in units of $h_{70}^{-2}10^{12}L_{\sun}$.
Table~\ref{tabfit} shows the main results.  Figure~ \ref{figlxlo} also
shows the fitted relations for the two alternative estimates of $L_{\rm
  X}$.  For each comparison of the above list, the $90\%$
c.l. ellipses overlap (see also
the inset plot in Fig.~ \ref{figlxlo}). The same result was obtained 
for $L_{\rm opt}(<0.5R_{200})$.

\input{tabfit}

\section{Discussion and conclusions}
\label{discu}

\begin{figure}
\centering 
\resizebox{\hsize}{!}{\includegraphics{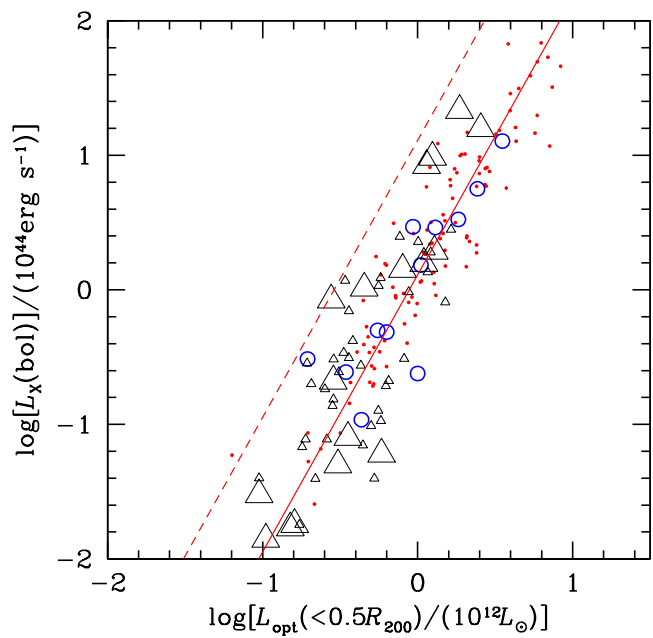}}
\caption
{Comparison with previous literature. CONF-FGSs and CLs as in
  Fig.~\ref{figlxlo}, but for the bolometric X-ray luminosity and the
  optical luminosity computed within $0.5R_{200}$.  The dashed line
  indicates overestimates by a factor of 10 in $L_{\rm X}{\rm (bol)}$ or
  underestimates by a factor of 3 in $L_{\rm opt}$ with respect to our
  CL sample (red points fitted by the solid line).  Large and small
  black triangles indicate the fossil and comparison systems in
  Harrison et al. (\cite{har12}, see their Fig.~5).  }
\label{figharrison}
\end{figure}

From the comparison between FGSs and CLs presented above 
we conclude that fossil systems are not significantly
  distinguishable from normal galaxy systems in the $L_{\rm
    opt}$--$L_{\rm X}$ plane.  In particular, we find no
evidence in favor of fossil systems being X-ray overluminous (by a
factor of $\sim$10, Khosroshahi et al. \cite{kho07}, see their Fig.~2) or
optically underluminous (by a factor of $\sim$3, Proctor et
al. \cite{pro11}, see their Sect.~5.3 and their Fig.~4) than normal
systems.  Differences such as those suggested in previous studies are
inconsistent with the plot shown in Fig.~\ref{figlxlo} (see also
Fig.~\ref{figharrison}), although there is still space to accommodate
modest differences. We plan some future efforts to reduce the
scatter of the S07 FGSs around the $L_{\rm X}$--$L_{\rm opt}$
relation, e.g.,  using FOGO redshift data, to further improve the
optical luminosity estimates. To improve the study of X-ray
properties, we have obtained X-ray Suzaku data for ten FGSs (data under
reduction).

Voevodkin et al. (\cite{voe10}) suggest that results might be dubious
when obtained for fossil and comparison systems treated in a
nonhomogeneous way. Voevodkin et al. (\cite{voe10}) present a
remarkable homogeneous comparison, finding no difference between
fossil and comparison systems. However, the fitted relation we
obtained for the CL sample, ${\rm log}[L_{\rm X,44}(0.5-2.0 {\rm
    keV})]=(-0.59\pm0.04)+(1.90\pm 0.09) \cdot {\rm log}[L_{\rm
    opt,12}<(R_{500})]$, is strongly inconsistent with their Fig.~5;
their optical luminosities are too small.  Since Proctor et
al. (\cite{pro11}) report the presence of a serious error in the
Voevodkin et al. (\cite{voe10}) estimations of the optical
luminosities, we no longer consider a quantitative comparison.
Rather, we consider the results of Harrison et al. (\cite{har12}), the
most recent paper on this subject and where the data were treated in a
homogeneous way.  We note that their X-ray luminosity estimates are
based on a different data source (XMM-{\em Newton}) and
methodology. Their estimation of optical luminosities is also quite
different, as it is based on galaxies in the red sequence and with no
(even modest) background subtraction and no LF extrapolation. In
addition, their estimation of characteristic radii is different.
Figure~\ref{figharrison} shows that the combination of data from
different sources can somehow generate difficulties when pursuing
precise comparisons. Thus, we stress the need to perform comparisons
based on a homogeneous treatment of the data.  Independently, both
Harrison et al. (\cite{har12}) and our study agree: there is no
difference between fossil and normal systems. These are the two most
recent studies on the $L_{\rm X}$--$L_{\rm opt}$ relation and treat a
total of about 30 fossil systems.  Thus, a definitive conclusion on
this issue has likely been reached.

Regarding the formation and evolution of fossil systems, our results
are consistent with the classical merging scenario where the large BCG
is the product of mergers/cannibalism of other group galaxies, with
the conservation of the galaxy optical light. In a general context,
Lin \& Mohr (\cite{lin04}) argue that BCGs grow in luminosity mainly by
merging with other luminous galaxies as the host clusters grow
hierarchically. The evidence that very luminous galaxies grow in
luminosity and decrease in number as the parent cluster evolves is the
result of a study based on merging vs. relaxed clusters (Barrena et
al. \cite{bar12}).  We propose that this process was particularly
efficient in fossil systems with the BCG growing at the expense of the
other brightest galaxies in the system.  Our Paper IV---using a
subsample of S07 FGSs for which we have computed $\sigma_v$---shows
that the main difference between fossil and normal systems of
comparable mass is the fraction of optical luminosity contributed by
the BCG.  On the other hand, it seems that the merging/coalescing
process does not cause any peculiarity in the global state of the hot
intracluster medium or, alternatively, that possible peculiarities are
a very short-lived phenomenon.

\begin{acknowledgements}

MG acknowledges financial support from the PRIN-INAF/2010 and MIUR
PRIN/2010-2011 (J91J12000450001).  ED gratefully acknowledges the
support of the Alfred P. Sloan Foundation.  JMA acknowledges support
from the European Research Council Starting Grant (SEDmorph;
P.I. V. Wild).  EMC is supported by Padua University (grants
60A02-1283/10, 5052/11, 4807/12).  JIP and JVM acknowledge financial
support from the Spanish MINECO under grant AYA2010-21887-C04-01, and
from Junta de Andaluc\'{\i}a Excellence Project PEX2011-FQM7058.  This
work has been supported by the Programa Nacional de Astronom\'ia y
Astrof\'isica of the Spanish Ministry of Science and Innovation under
grants AYA2010-21887-C04-04, AYA2007-67965-C03-01 and under the
Consolider-Ingenio 2010 Program grant CSD2006-00070: First Science
with the GTC (http://www.iac.es/consolider-ingenio-gtc).  This work
was partially funded by the local Canarian Government (grant
ProID20100140).  This article is based on observations made with the
Telescopio Nazionale Galileo operated on the island of La Palma by the
''Fundaci\'on Galileo Galilei-INAF, Fundaci\'on Canaria,'' in the
Spanish Observatorio del Roque de los Muchachos of the Instituto de
Astrof\'isica de Canarias.  This research has made use of the
NASA/IPAC Extragalactic Database (NED) which is operated by the Jet
Propulsion Laboratory, California Institute of Technology, under
contract with the National Aeronautics and Space Administration.
Funding for the SDSS and SDSS-II has been provided by the Alfred
P. Sloan Foundation, the Participating Institutions, the National
Science Foundation, the U.S. Department of Energy, the National
Aeronautics and Space Administration, the Japanese Monbukagakusho, the
Max Planck Society, and the Higher Education Funding Council for
England. The SDSS Web Site is http://www.sdss.org/.  The SDSS is
managed by the Astrophysical Research Consortium for the Participating
Institutions. The Participating Institutions are the American Museum
of Natural History, Astrophysical Institute Potsdam, University of
Basel, University of Cambridge, Case Western Reserve University,
University of Chicago, Drexel University, Fermilab, the Institute for
Advanced Study, the Japan Participation Group, Johns Hopkins
University, the Joint Institute for Nuclear Astrophysics, the Kavli
Institute for Particle Astrophysics and Cosmology, the Korean
Scientist Group, the Chinese Academy of Sciences (LAMOST), Los Alamos
National Laboratory, the Max-Planck-Institute for Astronomy (MPIA),
the Max-Planck-Institute for Astrophysics (MPA), New Mexico State
University, Ohio State University, University of Pittsburgh,
University of Portsmouth, Princeton University, the United States
Naval Observatory, and the University of Washington.

\end{acknowledgements}

\end{document}

%% file: tabfg.tex
%new commands
%\def\lesssim{\mathrel{\hbox{\rlap{\hbox{\lower4pt\hbox{$\sim$}}}\hbox{$<$}}}}
%\def\gtrsim{\mathrel{\hbox{\rlap{\hbox{\lower4pt\hbox{$\sim$}}}\hbox{$>$}}}}
%\newcommand{\mincir}{\raise -2.truept\hbox{\rlap{\hbox{$\sim$}}\raise5.truept
%\hbox{$<$}\ }}
%\newcommand{\magcir}{\raise -2.truept\hbox{\rlap{\hbox{$\sim$}}\raise5.truept
%\hbox{$>$}\ }}
%\newcommand{\siml}{\raise -2.truept\hbox{\rlap{\hbox{$\sim$}}\raise5.truept
%\hbox{$<$}\ }}
%\newcommand{\simg}{\raise -2.truept\hbox{\rlap{\hbox{$\sim$}}\raise5.truept
%\hbox{$>$}\ }}
%\newcommand{\be}{\begin{equation}}
%\newcommand{\ee}{\end{equation}}
%\newcommand{\ba}{\begin{eqnarray}}
%\newcommand{\ea}{\end{eqnarray}}
%\newcommand {\h} {$h^{-1}$ Mpc $ \;$}
%\newcommand {\kpc} {$h^{-1}$ kpc}
%\newcommand {\hh} {$h^{-1}$ Mpc}
%\newcommand {\ks} {km~s$^{-1} \;$}
%\newcommand {\kss} {km~s$^{-1}$}
%\newcommand {\mpc} {$Mpc \;$}
%\newcommand {\msun} {$h^{-1} \  M_{\odot} \;$}
%\newcommand {\m} {$M_{\odot} \;$}
%\newcommand {\ml} {$h \, M_{\odot}/L_{\odot} \;$}
%\newcommand {\mll} {$h \, M_{\odot}/L_{\odot}$}
%\newcommand{\vel}{\,{\rm km\,s^{-1}}}
%\newcommand{\tng}{\mathrm{T}}
%\newcommand{\sds}{\mathrm{S}}
%\newcommand{\tns}{\mathrm{T+S}}
%%
%\addtocounter{table}{-2}
\begin{table*}[!ht]
%\begin{table}[!ht]
        \caption[]{Properties of the FGS sample.}
         \label{tabfg}
              $$ 
        % \begin{array}{p{0.5\linewidth}l}
           \begin{array}{l c c c c c c c  l }
            \hline
            \noalign{\smallskip}
            \hline
            \noalign{\smallskip}
\mathrm{ID} &\mathrm{Notes}&\mathrm{\alpha},\mathrm{\delta}\,(\mathrm{J}2000)  & z& L_{\rm X}{\rm (0.1-2.4)\,keV}&R_{500}&L_{\rm opt}(<R_{500})&L_{\rm opt}(<0.5R_{200})&{\rm Other\ catalogs}\\
&  & & & h_{70}^{-2}{\rm erg/s^{-1}}&h_{70}^{-1}{\rm Mpc}& h_{70}^{-2}L_{\sun}& h_{70}^{-2}L_{\sun}&{\rm References}\\
            \hline
            \noalign{\smallskip}  
{\mathrm {FGS}}01&{\mathrm d}& 01\ 50\ 21.30  ,-10\ 05\ 30.5&   0.365& 4.87{\rm E+44}&1.08& 2.50{\rm E+12}&  2.00{\rm E+12}  &8\\
{\mathrm {FGS}}02&{\mathrm c}& 01\ 52\ 42.00  ,+01\ 00\ 25.6&   0.230& 5.21{\rm E+44}&1.19& 5.02{\rm E+12}&  3.52{\rm E+12}  &2 {\rm \ (\object {Abell\ 267})}\\
{\mathrm {FGS}}03&{\mathrm c}& 07\ 52\ 44.20  ,+45\ 56\ 57.4&   0.052& 2.21{\rm E+43}&0.63& 3.05{\rm E+11}&  1.95{\rm E+11}  &-\\
{\mathrm {FGS}}04&{\mathrm d}& 08\ 07\ 30.80  ,+34\ 00\ 41.6&   0.208& 1.71{\rm E+44}&0.93& 1.22{\rm E+12}&  1.06{\rm E+12}  &5\\
{\mathrm {FGS}}05&{\mathrm d}& 08\ 42\ 57.60  ,+36\ 21\ 59.3&   0.282& 1.02{\rm E+45}&1.35& 4.89{\rm E+12}&  3.73{\rm E+12}  &2 {\rm \ (\object {Abell\ 697})}\\
{\mathrm {FGS}}06&{\mathrm d}& 08\ 44\ 56.60  ,+42\ 58\ 35.7&   0.054& 0.66{\rm E+43}&0.48& 3.89{\rm E+11}&  3.15{\rm E+11}  &-\\
{\mathrm {FGS}}07&{\mathrm d}& 09\ 03\ 03.20  ,+27\ 39\ 29.4&   0.489& 5.88{\rm E+44}&1.05& 3.06{\rm E+12}&  2.84{\rm E+12}  &6, 7\\
{\mathrm {FGS}}08&{\mathrm c}& 09\ 48\ 29.00  ,+49\ 55\ 06.7&   0.409& 1.63{\rm E+44}&0.82& 3.56{\rm E+12}&  1.30{\rm E+12}  &-\\
{\mathrm {FGS}}09&{\mathrm d}& 10\ 43\ 02.60  ,+00\ 54\ 18.3&   0.125& 1.98{\rm E+44}&1.01& 1.51{\rm E+12}&  1.03{\rm E+12}  &4\\
{\mathrm {FGS}}10&{\mathrm c}& 10\ 54\ 52.00  ,+55\ 21\ 12.5&   0.468& 2.80{\rm E+44}&0.90& 3.28{\rm E+12}&  2.43{\rm E+12}  &7\\
{\mathrm {FGS}}11&{\mathrm a}& 11\ 14\ 39.80  ,+40\ 37\ 35.2&   0.202& 1.21{\rm E+44}&0.86&-        &    -                   &-\\        
{\mathrm {FGS}}12&{\mathrm d}& 11\ 21\ 55.30  ,+10\ 49\ 23.2&   0.240& 1.32{\rm E+44}&0.86& 3.22{\rm E+11}&  3.79{\rm E+11}  &5,8\\
{\mathrm {FGS}}13&{\mathrm d}& 11\ 41\ 28.30  ,+05\ 58\ 29.5&   0.188& 7.18{\rm E+43}&0.77& 1.62{\rm E+12}&  1.12{\rm E+12}  &-\\
{\mathrm {FGS}}14&{\mathrm c}& 11\ 46\ 47.60  ,+09\ 52\ 28.2&   0.221& 1.78{\rm E+44}&0.93& 1.97{\rm E+12}&  1.83{\rm E+12}  &1\\
{\mathrm {FGS}}15&{\mathrm a}& 11\ 48\ 03.80  ,+56\ 54\ 25.6&   0.105& 2.67{\rm E+43}&0.64&-        &    -                   &-\\       
{\mathrm {FGS}}16&{\mathrm d}& 11\ 49\ 15.00  ,+48\ 11\ 04.9&   0.283& 2.01{\rm E+44}&0.93& 2.08{\rm E+12}&  1.48{\rm E+12}  &1, 5, 7\\
{\mathrm {FGS}}17&{\mathrm c}& 12\ 47\ 42.10  ,+41\ 31\ 37.7&   0.155& 1.80{\rm E+43}&0.57& 2.93{\rm E+11}&  3.45{\rm E+11}  &5, 6\\
{\mathrm {FGS}}18&{\mathrm d}& 13\ 00\ 09.40  ,+44\ 43\ 01.3&   0.233& 7.41{\rm E+43}&0.76& 7.09{\rm E+11}&  5.96{\rm E+11}  &-\\
{\mathrm {FGS}}19&{\mathrm b}& 13\ 35\ 60.00  ,-03\ 31\ 29.2&   0.177& 1.28{\rm E+44}&0.89& -       &    -      -            &1, 5\\      
{\mathrm {FGS}}20&{\mathrm c}& 14\ 10\ 04.20  ,+41\ 45\ 20.9&   0.094& 0.80{\rm E+43}&0.49& 4.84{\rm E+11}&  4.34{\rm E+11}  &3, 4, 5\\
{\mathrm {FGS}}21&{\mathrm d}& 14\ 45\ 16.90  ,+00\ 39\ 34.3&   0.306& 2.70{\rm E+44}&0.98& 1.54{\rm E+12}&  1.50{\rm E+12}  &8\\
{\mathrm {FGS}}22&{\mathrm d}& 14\ 53\ 59.00  ,+48\ 24\ 17.1&   0.146& 1.77{\rm E+43}&0.57& 5.50{\rm E+11}&  4.68{\rm E+11}  &5\\
{\mathrm {FGS}}23&{\mathrm c}& 15\ 29\ 46.30  ,+44\ 08\ 04.2&   0.148& 3.49{\rm E+43}&0.67& 8.06{\rm E+11}&  5.52{\rm E+11}  &5, 8\\
{\mathrm {FGS}}24&{\mathrm d}& 15\ 33\ 44.10  ,+03\ 36\ 57.5&   0.293& 2.32{\rm E+44}&0.95& 2.61{\rm E+12}&  1.81{\rm E+12}  &5, 8\\
{\mathrm {FGS}}25&{\mathrm d}& 15\ 39\ 50.80  ,+30\ 43\ 04.0&   0.097& 1.67{\rm E+44}&0.98& 1.71{\rm E+12}&  1.29{\rm E+12}  &2 {\rm \ (\object {Abell\ 2110})}\\
{\mathrm {FGS}}26&{\mathrm c}& 15\ 48\ 55.90  ,+08\ 50\ 44.4&   0.072& 1.75{\rm E+43}&0.59& 1.33{\rm E+12}&  1.00{\rm E+12}  &3\\
{\mathrm {FGS}}27&{\mathrm c}& 16\ 14\ 31.10  ,+26\ 43\ 50.4&   0.184& 9.24{\rm E+43}&0.82& 1.34{\rm E+12}&  1.05{\rm E+12}  &8\\
{\mathrm {FGS}}28&{\mathrm a}& 16\ 37\ 20.50  ,+41\ 11\ 20.3&   0.032& 0.09{\rm E+43}&0.31&-        &    -                   &-\\       
{\mathrm {FGS}}29&{\mathrm a}& 16\ 47\ 02.10  ,+38\ 50\ 04.3&   0.135& 1.93{\rm E+43}&0.59&-         &   -                   &-\\        
{\mathrm {FGS}}30&{\mathrm c}& 17\ 18\ 11.90  ,+56\ 39\ 56.1&   0.114& 1.58{\rm E+44}&0.96& 1.08{\rm E+12}&  9.34{\rm E+11}  &7, 8\\
{\mathrm {FGS}}31&{\mathrm d}& 17\ 20\ 10.00  ,+26\ 37\ 32.1&   0.159& 6.68{\rm E+44}&1.31& 3.81{\rm E+12}&  2.68{\rm E+12}  &7, 8\\
{\mathrm {FGS}}32&{\mathrm a}& 17\ 28\ 52.20  ,+55\ 16\ 40.8&   0.148& 1.15{\rm E+43}&0.52&-        &    -                   &-\\        
{\mathrm {FGS}}33&{\mathrm d}& 22\ 56\ 30.00  ,-00\ 32\ 10.7&   0.224& 9.16{\rm E+43}&0.80& 3.03{\rm E+12}&  2.56{\rm E+12}  &7, 8\\
{\mathrm {FGS}}34&{\mathrm c}& 23\ 58\ 15.10  ,+15\ 05\ 43.6&   0.178& 3.41{\rm E+43}&0.66& 7.03{\rm E+11}&  6.31{\rm E+11}  &-\\
                        \noalign{\smallskip}			    
            \hline					    
            \noalign{\smallskip}			    
            \hline					    
         \end{array}
     $$ 
\tablefoot{(a)~With no clear corrisponding density peak in the 2D galaxy distribution (see Sect.~\ref{2d}); (b)~not fully sampled by SDSS-DR7; (c)~with confirmed fossil classification according to Paper III (our CONF-{FGS} sample); (d)~our NOCONF-{FGS} sample.}
\tablebib{
(1)~Zwicky \& Kowal (\cite{zwi68}) and catalogs therein; (2)~Abell et al. (\cite{abe89}, Abell-ACO); (3)~Gal et al. (\cite{gal03}, NSC Northern Sky Optical Cluster Survey); (4)~Miller et al. (\cite{mil05}, SDSS-C4); Koester et al. (\cite{koe07}, MaxBCG); McConnachie et al. (\cite{mcc09}, SDSSCGB); Wen et al. (\cite{wen09}, \cite{wen10}, WHL); Hao et al. (\cite{hao10}, GMBCG). For each system, the list is not meant to be exhaustive (see NED for this).
}

\end{table*}
%\end{table}

%%

%% file: tabpop1.tex
%new commands
%\def\lesssim{\mathrel{\hbox{\rlap{\hbox{\lower4pt\hbox{$\sim$}}}\hbox{$<$}}}}
%\def\gtrsim{\mathrel{\hbox{\rlap{\hbox{\lower4pt\hbox{$\sim$}}}\hbox{$>$}}}}
%\newcommand{\mincir}{\raise -2.truept\hbox{\rlap{\hbox{$\sim$}}\raise5.truept
%\hbox{$<$}\ }}
%\newcommand{\magcir}{\raise -2.truept\hbox{\rlap{\hbox{$\sim$}}\raise5.truept
%\hbox{$>$}\ }}
%\newcommand{\siml}{\raise -2.truept\hbox{\rlap{\hbox{$\sim$}}\raise5.truept
%\hbox{$<$}\ }}
%\newcommand{\simg}{\raise -2.truept\hbox{\rlap{\hbox{$\sim$}}\raise5.truept
%\hbox{$>$}\ }}
%\newcommand{\be}{\begin{equation}}
%\newcommand{\ee}{\end{equation}}
%\newcommand{\ba}{\begin{eqnarray}}
%\newcommand{\ea}{\end{eqnarray}}
%\newcommand {\h} {$h^{-1}$ Mpc $ \;$}
%\newcommand {\kpc} {$h^{-1}$ kpc}
%\newcommand {\hh} {$h^{-1}$ Mpc}
%\newcommand {\ks} {km~s$^{-1} \;$}
%\newcommand {\kss} {km~s$^{-1}$}
%\newcommand {\mpc} {$Mpc \;$}
%\newcommand {\msun} {$h^{-1} \  M_{\odot} \;$}
%\newcommand {\m} {$M_{\odot} \;$}
%\newcommand {\ml} {$h \, M_{\odot}/L_{\odot} \;$}
%\newcommand {\mll} {$h \, M_{\odot}/L_{\odot}$}
%\newcommand{\vel}{\,{\rm km\,s^{-1}}}
%\newcommand{\tng}{\mathrm{T}}
%\newcommand{\sds}{\mathrm{S}}
%\newcommand{\tns}{\mathrm{T+S}}
%%
%\addtocounter{table}{-2}
\begin{table*}[!ht]
%\begin{table}[!ht]
        \caption[]{Properties of the CL sample.}
         \label{tabpop1}
              $$ 
        % \begin{array}{p{0.5\linewidth}l}
           \begin{array}{l c c c c c c }
            \hline
            \noalign{\smallskip}
            \hline
            \noalign{\smallskip}
\mathrm{ID} &\mathrm{\alpha},\mathrm{\delta}\,(\mathrm{J}2000)  & z& L_{\rm X}{\rm (0.1-2.4)\,keV}&R_{500}&L_{\rm opt}(<R_{500})&L_{\rm opt}(<0.5R_{200})\\
  & & & h_{70}^{-2}{\rm erg/s^{-1}}&h_{70}^{-1}{\rm Mpc}& h_{70}^{-2}L_{\sun} &h_{70}^{-2}L_{\sun}\\
            \hline
            \noalign{\smallskip}  
{\mathrm {CL}}001&   00\ 41\ 50.09  ,-09\ 18\ 06.8&   0.052&   4.24{\rm  E+44}&  1.24&2.81{\rm E+12} & 2.02{\rm E+}12\\                
{\mathrm {CL}}002&   01\ 14\ 56.40  ,+00\ 22\ 28.6&   0.047&   4.34{\rm  E+43}&  0.74&1.36{\rm E+12} & 9.64{\rm E+11}\\
{\mathrm {CL}}003&   01\ 19\ 37.73  ,+14\ 53\ 35.2&   0.129&   1.29{\rm  E+44}&  0.91&3.08{\rm E+12} & 2.16{\rm E+12}\\
{\mathrm {CL}}004&   01\ 37\ 15.36  ,-09\ 12\ 10.1&   0.039&   2.44{\rm  E+43}&  0.65&5.70{\rm E+11} & 4.88{\rm E+11}\\
{\mathrm {CL}}006&   07\ 36\ 24.96  ,+39\ 25\ 58.4&   0.117&   2.75{\rm  E+44}&  1.09&1.39{\rm E+12} & 1.14{\rm E+12}\\
{\mathrm {CL}}007&   07\ 47\ 00.89  ,+41\ 31\ 53.0&   0.028&   4.21{\rm  E+42}&  0.44&1.31{\rm E+11} & 6.35{\rm E+10}\\
{\mathrm {CL}}008&   07\ 53\ 18.98  ,+29\ 22\ 26.8&   0.062&   5.65{\rm  E+43}&  0.78&1.17{\rm E+12} & 7.23{\rm E+11}\\
{\mathrm {CL}}009&   07\ 58\ 28.13  ,+37\ 47\ 19.7&   0.041&   1.60{\rm  E+42}&  0.35&2.36{\rm E+11} & 2.16{\rm E+11}\\
{\mathrm {CL}}010&   08\ 00\ 58.68  ,+36\ 02\ 48.8&   0.288&   5.77{\rm  E+44}&  1.18&4.39{\rm E+12} & 3.20{\rm E+12 }\\
{\mathrm {CL}}011&   08\ 09\ 40.25  ,+34\ 55\ 34.3&   0.080&   7.91{\rm  E+43}&  0.83&7.03{\rm E+11} & 5.80{\rm E+11 }\\
{\mathrm {CL}}012&   08\ 10\ 22.61  ,+42\ 16\ 00.8&   0.064&   2.78{\rm  E+43}&  0.66&6.75{\rm E+11} & 4.50{\rm E+11 }\\
{\mathrm {CL}}013&   08\ 22\ 10.01  ,+47\ 05\ 58.2&   0.130&   3.03{\rm  E+44}&  1.11&2.53{\rm E+12} & 1.62{\rm E+12}\\
{\mathrm {CL}}014&   08\ 24\ 05.02  ,+03\ 26\ 17.9&   0.347&   1.06{\rm  E+44}&  0.77&1.38{\rm E+12} & 5.39{\rm E+11}\\
{\mathrm {CL}}015&   08\ 25\ 27.65  ,+47\ 07\ 10.6&   0.126&   2.83{\rm  E+44}&  1.09&3.75{\rm E+12} & 2.72{\rm E+12}\\
{\mathrm {CL}}016&   08\ 28\ 06.67  ,+44\ 45\ 48.2&   0.145&   2.37{\rm  E+44}&  1.04&2.23{\rm E+12} & 1.68{\rm E+12}\\
{\mathrm {CL}}019&   08\ 50\ 11.98  ,+36\ 03\ 41.0&   0.373&   1.09{\rm  E+45}&  1.30&9.64{\rm E+12} & 7.38{\rm E+12}\\
{\mathrm {CL}}020&   09\ 13\ 45.86  ,+40\ 56\ 02.0&   0.442&   1.01{\rm  E+45}&  1.23&3.42{\rm E+12} & 3.96{\rm E+12}\\
{\mathrm {CL}}021&   09\ 13\ 46.70  ,+47\ 42\ 07.6&   0.051&   3.66{\rm  E+43}&  0.71&5.31{\rm E+11} & 4.68{\rm E+11}\\
{\mathrm {CL}}022&   09\ 17\ 51.29  ,+51\ 43\ 20.3&   0.217&   7.35{\rm  E+44}&  1.29&6.71{\rm E+12} & 5.90{\rm E+12}\\
{\mathrm {CL}}023&   09\ 43\ 02.40  ,+47\ 00\ 13.7&   0.406&   4.98{\rm  E+44}&  1.06&8.90{\rm E+12} & 7.11{\rm E+12}\\
{\mathrm {CL}}024&   09\ 47\ 08.69  ,+54\ 28\ 31.4&   0.046&   2.46{\rm  E+43}&  0.65&5.23{\rm E+11} & 4.07{\rm E+11}\\
{\mathrm {CL}}025&   09\ 52\ 48.22  ,+51\ 53\ 19.7&   0.214&   5.03{\rm  E+44}&  1.19&2.18{\rm E+12} & 1.35{\rm E+12}\\
{\mathrm {CL}}026&   09\ 53\ 41.54  ,+01\ 42\ 42.5&   0.098&   5.45{\rm  E+43}&  0.76&5.71{\rm E+11} & 4.46{\rm E+11}\\
{\mathrm {CL}}027&   10\ 00\ 30.24  ,+44\ 09\ 18.0&   0.154&   1.67{\rm  E+44}&  0.95&7.90{\rm E+11} & 7.00{\rm E+11}\\
{\mathrm {CL}}028&   10\ 13\ 44.83  ,-00\ 06\ 30.6&   0.093&   7.00{\rm  E+43}&  0.81&1.34{\rm E+12} & 1.28{\rm E+12}\\
{\mathrm {CL}}029&   10\ 17\ 35.04  ,+59\ 33\ 27.7&   0.353&   1.44{\rm  E+45}&  1.40&1.01{\rm E+13} & 8.36{\rm E+12}\\
{\mathrm {CL}}030&   10\ 22\ 30.79  ,+50\ 06\ 10.8&   0.158&   3.41{\rm  E+44}&  1.12&3.40{\rm E+12} & 2.91{\rm E+12}\\
{\mathrm {CL}}031&   10\ 23\ 39.00  ,+04\ 11\ 14.3&   0.285&   1.90{\rm  E+45}&  1.55&4.98{\rm E+12} & 3.84{\rm E+12}\\
{\mathrm {CL}}032&   10\ 23\ 41.09  ,+49\ 08\ 05.6&   0.144&   4.23{\rm  E+44}&  1.18&2.71{\rm E+12} & 1.88{\rm E+12}\\
{\mathrm {CL}}033&   10\ 53\ 44.38  ,+54\ 52\ 21.4&   0.075&   5.28{\rm  E+43}&  0.76&1.12{\rm E+12} & 8.75{\rm E+11}\\
{\mathrm {CL}}034&   10\ 58\ 26.33  ,+56\ 47\ 31.9&   0.136&   3.54{\rm  E+44}&  1.14&3.22{\rm E+12} & 2.73{\rm E+12}\\
{\mathrm {CL}}035&   10\ 58\ 27.65  ,+01\ 34\ 05.5&   0.039&   1.06{\rm  E+43}&  0.54&6.65{\rm E+11} & 3.64{\rm E+11}\\
{\mathrm {CL}}036&   11\ 13\ 22.70  ,+02\ 32\ 32.6&   0.075&   1.07{\rm  E+44}&  0.90&1.67{\rm E+12} & 1.12{\rm E+12}\\
{\mathrm {CL}}037&   11\ 14\ 23.90  ,+58\ 23\ 26.5&   0.206&   3.64{\rm  E+44}&  1.11&1.29{\rm E+12} & 1.20{\rm E+12}\\
{\mathrm {CL}}038&   11\ 15\ 32.23  ,+54\ 26\ 05.6&   0.069&   3.83{\rm  E+43}&  0.71&1.07{\rm E+12} & 8.22{\rm E+11}\\
{\mathrm {CL}}039&   11\ 15\ 53.95  ,+01\ 29\ 44.2&   0.349&   1.62{\rm  E+45}&  1.44&1.06{\rm E+13} & 6.92{\rm E+12}\\
{\mathrm {CL}}040&   11\ 21\ 36.19  ,+48\ 03\ 50.0&   0.112&   9.05{\rm  E+43}&  0.85&2.04{\rm E+12} & 1.51{\rm E+12}\\
{\mathrm {CL}}041&   11\ 21\ 44.83  ,+02\ 48\ 51.5&   0.046&   2.84{\rm  E+43}&  0.67&9.56{\rm E+11} & 8.55{\rm E+11}\\
{\mathrm {CL}}042&   11\ 33\ 17.28  ,+66\ 22\ 45.5&   0.116&   1.51{\rm  E+44}&  0.95&1.63{\rm E+12} & 1.15{\rm E+}12\\
{\mathrm {CL}}043&   11\ 34\ 50.83  ,+49\ 03\ 46.4&   0.034&   1.96{\rm  E+43}&  0.62&7.17{\rm E+11} & 6.30{\rm E+11}\\
{\mathrm {CL}}045&   11\ 44\ 04.85  ,+05\ 48\ 11.2&   0.103&   7.28{\rm  E+43}&  0.81&1.21{\rm E+12} & 1.06{\rm E+12}\\
{\mathrm {CL}}046&   11\ 44\ 40.85  ,+67\ 24\ 40.0&   0.115&   1.72{\rm  E+44}&  0.98&1.84{\rm E+12} & 1.50{\rm E+12}\\
{\mathrm {CL}}047&   11\ 59\ 17.50  ,+49\ 47\ 46.3&   0.210&   3.39{\rm  E+44}&  1.09&3.01{\rm E+12} & 1.79{\rm E+12}\\
{\mathrm {CL}}048&   12\ 00\ 24.48  ,+03\ 19\ 51.6&   0.133&   3.94{\rm  E+44}&  1.17&3.60{\rm E+12} & 2.49{\rm E+12}\\
{\mathrm {CL}}049&   12\ 04\ 25.18  ,+01\ 54\ 01.8&   0.020&   1.51{\rm  E+43}&  0.59&5.58{\rm E+11} & 4.91{\rm E+11}\\
{\mathrm {CL}}051&   12\ 17\ 40.80  ,+03\ 39\ 41.0&   0.076&   2.75{\rm  E+44}&  1.11&2.86{\rm E+12} & 2.40{\rm E+12}\\
{\mathrm {CL}}053&   12\ 27\ 50.28  ,+63\ 23\ 01.3&   0.145&   1.26{\rm  E+44}&  0.90&1.54{\rm E+12} & 1.24{\rm E+12}\\
{\mathrm {CL}}054&   12\ 36\ 59.18  ,+63\ 11\ 29.0&   0.301&   5.87{\rm  E+44}&  1.17&7.20{\rm E+12} & 5.71{\rm E+12}\\
{\mathrm {CL}}056&   12\ 47\ 43.20  ,-02\ 47\ 31.6&   0.179&   2.80{\rm  E+44}&  1.06&3.37{\rm E+12} & 2.77{\rm E+12}\\
{\mathrm {CL}}057&   12\ 58\ 41.09  ,-01\ 45\ 24.8&   0.084&   3.48{\rm  E+44}&  1.17&2.45{\rm E+12} & 1.73{\rm E+12}\\
{\mathrm {CL}}058&   13\ 02\ 50.69  ,-02\ 30\ 22.3&   0.083&   6.04{\rm  E+43}&  0.78&1.18{\rm E+12} & 8.11{\rm E+11}\\
{\mathrm {CL}}059&   13\ 03\ 56.50  ,+67\ 31\ 03.7&   0.106&   1.98{\rm  E+43}&  0.60&8.17{\rm E+11} & 5.05{\rm E+11}\\
{\mathrm {CL}}060&   13\ 09\ 16.99  ,-01\ 36\ 45.4&   0.088&   9.30{\rm  E+43}&  0.86&7.52{\rm E+11} & 6.16{\rm E+11}\\
{\mathrm {CL}}061&   13\ 11\ 30.00  ,-01\ 20\ 07.4&   0.181&   1.23{\rm  E+45}&  1.48&6.78{\rm E+12} & 5.34{\rm E+12}\\
{\mathrm {CL}}062&   13\ 14\ 22.85  ,+64\ 34\ 44.0&   0.220&   4.35{\rm  E+44}&  1.15&2.47{\rm E+12} & 1.92{\rm E+12}\\
{\mathrm {CL}}063&   13\ 25\ 49.99  ,+59\ 19\ 20.6&   0.151&   1.88{\rm  E+44}&  0.98&1.75{\rm E+12} & 1.46{\rm E+12}\\
{\mathrm {CL}}064&   13\ 26\ 17.83  ,+00\ 13\ 32.5&   0.082&   9.14{\rm  E+43}&  0.86&7.58{\rm E+11} & 6.29{\rm E+11}\\
                                       \noalign{\smallskip}			    
            \hline					    
            \noalign{\smallskip}			    
            \hline					    
         \end{array}
     $$        
         \end{table*}
\addtocounter{table}{-1}
\begin{table*}[!ht]
          \caption[ ]{Continued.}
     $$        
           \begin{array}{l c c c  c c c c }
            \hline
            \noalign{\smallskip}
            \hline
            \noalign{\smallskip}
\mathrm{ID} &\mathrm{\alpha},\mathrm{\delta}\,(\mathrm{J}2000)  & z& L_{\rm X}{\rm (0.1-2.4)\,keV}&R_{500}&L_{\rm opt}(<R_{500})&L_{\rm opt}(<0.5R_{200})\\
  & & & {\rm erg/s^{-1}} h_{70}^{-2}&{\rm Mpc}\,h_{70}^{-1}& L_{\sun} h_{70}^{-2}&L_{\sun} h_{70}^{-2}\\
            \hline
            \noalign{\smallskip}  
{\mathrm {CL}}065&   13\ 27\ 05.06  ,+02\ 11\ 53.5&   0.259&   5.24{\rm  E+44}&  1.17&5.31{\rm E+12} & 4.32{\rm E+12}\\
{\mathrm {CL}}066&   13\ 30\ 49.94  ,-01\ 52\ 22.1&   0.086&   1.13{\rm  E+44}&  0.90&2.07{\rm E+12} & 1.50{\rm E+12 }\\
{\mathrm {CL}}067&   13\ 32\ 38.90  ,+54\ 19\ 09.5&   0.101&   6.65{\rm  E+43}&  0.79&8.71{\rm E+11} & 6.92{\rm E+11}\\
{\mathrm {CL}}068&   13\ 36\ 06.53  ,+59\ 12\ 26.6&   0.070&   1.43{\rm  E+44}&  0.96&1.79{\rm E+12} & 1.46{\rm E+12}\\
{\mathrm {CL}}069&   13\ 42\ 05.47  ,+02\ 13\ 39.0&   0.077&   8.22{\rm  E+43}&  0.84&1.69{\rm E+12} & 1.27{\rm E+12}\\
{\mathrm {CL}}071&   13\ 53\ 00.77  ,+05\ 09\ 21.2&   0.079&   1.09{\rm  E+44}&  0.90&2.73{\rm E+12}&  2.40{\rm E+12} \\
{\mathrm {CL}}072&   13\ 59\ 53.14  ,+62\ 31\ 19.6&   0.329&   6.09{\rm  E+44}&  1.16&5.97{\rm E+12}&  3.37{\rm E+12} \\
{\mathrm {CL}}073&   14\ 01\ 02.45  ,+02\ 52\ 47.3&   0.252&   1.92{\rm  E+45}&  1.58&8.19{\rm E+12}&  6.26{\rm E+12} \\
{\mathrm {CL}}074&   14\ 11\ 24.07  ,+52\ 12\ 36.4&   0.460&   6.03{\rm  E+44}&  1.08&2.50{\rm E+12}&  2.09{\rm E+12} \\
{\mathrm {CL}}075&   14\ 15\ 14.21  ,-00\ 30\ 03.6&   0.136&   1.34{\rm  E+44}&  0.92&1.78{\rm E+12}&  1.40{\rm E+12} \\
{\mathrm {CL}}076&   14\ 24\ 48.48  ,+02\ 40\ 55.9&   0.052&   1.51{\rm  E+43}&  0.58&3.97{\rm E+11}&  3.69{\rm E+11} \\
{\mathrm {CL}}077&   14\ 25\ 22.92  ,+63\ 11\ 22.6&   0.139&   2.80{\rm  E+44}&  1.08&2.21{\rm E+12}&  1.63{\rm E+12} \\
{\mathrm {CL}}078&   14\ 28\ 51.31  ,+01\ 45\ 36.4&   0.320&   1.39{\rm  E+44}&  0.84&2.61{\rm E+12}&  2.07{\rm E+12} \\
{\mathrm {CL}}079&   14\ 38\ 25.27  ,+03\ 38\ 37.0&   0.224&   9.10{\rm  E+43}&  0.80&2.48{\rm E+12}&  2.06{\rm E+12} \\
{\mathrm {CL}}080&   14\ 40\ 38.47  ,+03\ 28\ 19.9&   0.027&   1.89{\rm  E+43}&  0.62&6.31{\rm E+11}&  5.26{\rm E+11} \\
{\mathrm {CL}}081&   14\ 52\ 55.01  ,+58\ 02\ 58.6&   0.317&   7.92{\rm  E+44}&  1.24&4.59{\rm E+12}&  3.95{\rm E+12} \\
{\mathrm {CL}}084&   15\ 11\ 33.53  ,+01\ 45\ 51.1&   0.037&   3.71{\rm  E+42}&  0.42&2.57{\rm E+11}&  1.97{\rm E+11} \\
{\mathrm {CL}}085&   15\ 12\ 51.05  ,-01\ 28\ 47.3&   0.122&   1.24{\rm  E+44}&  0.91&1.54{\rm E+12}&  1.20{\rm E+12} \\
{\mathrm {CL}}086&   15\ 16\ 19.18  ,+00\ 05\ 52.1&   0.118&   1.68{\rm  E+44}&  0.97&2.05{\rm E+12}&  1.75{\rm E+12} \\
{\mathrm {CL}}087&   15\ 16\ 34.03  ,-00\ 56\ 55.7&   0.115&   5.76{\rm  E+43}&  0.76&1.31{\rm E+12}&  9.19{\rm E+11} \\
{\mathrm {CL}}088&   15\ 29\ 12.05  ,+52\ 50\ 39.8&   0.072&   3.13{\rm  E+43}&  0.68&6.85{\rm E+11}&  4.79{\rm E+11} \\
{\mathrm {CL}}089&   15\ 44\ 29.81  ,+51\ 27\ 45.0&   0.158&   1.69{\rm  E+44}&  0.95&1.69{\rm E+12}&  1.22{\rm E+12} \\
{\mathrm {CL}}090&   16\ 01\ 22.13  ,+53\ 54\ 19.1&   0.106&   1.22{\rm  E+44}&  0.91&2.85{\rm E+12}&  2.40{\rm E+12} \\
{\mathrm {CL}}091&   16\ 11\ 17.69  ,+36\ 57\ 38.2&   0.067&   2.95{\rm  E+43}&  0.67&7.72{\rm E+11}&  6.12{\rm E+11} \\
{\mathrm {CL}}092&   16\ 17\ 33.00  ,+34\ 57\ 49.3&   0.030&   1.47{\rm  E+43}&  0.58&6.11{\rm E+11}&  5.21{\rm E+11} \\
{\mathrm {CL}}093&   16\ 27\ 40.13  ,+40\ 55\ 14.9&   0.030&   6.32{\rm  E+42}&  0.48&3.46{\rm E+11}&  3.17{\rm E+11} \\
{\mathrm {CL}}094&   16\ 27\ 24.41  ,+42\ 40\ 42.6&   0.031&   6.33{\rm  E+42}&  0.48&2.49{\rm E+11}&  1.96{\rm E+11} \\
{\mathrm {CL}}095&   16\ 29\ 41.88  ,+40\ 49\ 23.2&   0.031&   1.42{\rm  E+43}&  0.58&7.97{\rm E+11}&  5.14{\rm E+11} \\
{\mathrm {CL}}096&   16\ 40\ 22.10  ,+46\ 42\ 19.8&   0.228&   1.50{\rm  E+45}&  1.51&8.58{\rm E+12}&  5.92{\rm E+12} \\
{\mathrm {CL}}097&   16\ 54\ 44.47  ,+40\ 02\ 51.4&   0.100&   5.85{\rm  E+43}&  0.77&9.34{\rm E+11}&  7.21{\rm E+11} \\
{\mathrm {CL}}098&   16\ 56\ 20.28  ,+39\ 16\ 59.9&   0.061&   2.66{\rm  E+43}&  0.66&7.16{\rm E+11}&  5.23{\rm E+11} \\
{\mathrm {CL}}099&   16\ 59\ 45.36  ,+32\ 36\ 58.0&   0.101&   1.10{\rm  E+44}&  0.89&1.51{\rm E+12}&  1.09{\rm E+12} \\
{\mathrm {CL}}100&   17\ 02\ 42.62  ,+34\ 03\ 40.7&   0.095&   4.10{\rm  E+44}&  1.21&3.14{\rm E+12}&  2.51{\rm E+12} \\
{\mathrm {CL}}101&   17\ 12\ 47.62  ,+64\ 03\ 47.5&   0.080&   2.69{\rm  E+44}&  1.10&4.40{\rm E+12}&  3.73{\rm E+12} \\
{\mathrm {CL}}102&   17\ 15\ 21.60  ,+57\ 24\ 30.2&   0.028&   2.47{\rm  E+43}&  0.66&6.30{\rm E+11}&  5.69{\rm E+11} \\
{\mathrm {CL}}104&   17\ 20\ 09.22  ,+27\ 40\ 08.8&   0.164&   3.60{\rm  E+44}&  1.13&3.33{\rm E+12}&  2.80{\rm E+12} \\
{\mathrm {CL}}106&   21\ 25\ 12.38  ,-06\ 57\ 55.8&   0.115&   7.19{\rm  E+43}&  0.80&1.42{\rm E+12}&  1.04{\rm E+12} \\
{\mathrm {CL}}107&   21\ 29\ 40.54  ,+00\ 05\ 47.4&   0.234&   1.05{\rm  E+45}&  1.39&6.86{\rm E+12}&  4.49{\rm E+12} \\
{\mathrm {CL}}108&   21\ 55\ 40.54  ,+12\ 31\ 55.2&   0.192&   3.35{\rm  E+44}&  1.10&3.06{\rm E+12}&  2.50{\rm E+12} \\
{\mathrm {CL}}109&   21\ 57\ 25.75  ,-07\ 47\ 40.6&   0.061&   5.86{\rm  E+43}&  0.79&1.84{\rm E+12}&  1.31{\rm E+12} \\
{\mathrm {CL}}110&   22\ 14\ 49.82  ,+13\ 49\ 49.4&   0.025&   4.72{\rm  E+42}&  0.45&2.81{\rm E+11}&  2.37{\rm E+11} \\
{\mathrm {CL}}111&   22\ 16\ 15.48  ,-09\ 20\ 23.6&   0.082&   1.43{\rm  E+44}&  0.95&1.69{\rm E+12}&  9.40{\rm E+11} \\
{\mathrm {CL}}112&   23\ 24\ 21.05  ,+14\ 39\ 52.2&   0.042&   5.17{\rm  E+43}&  0.77&1.35{\rm E+12}&  9.92{\rm E+11} \\
{\mathrm {CL}}113&   23\ 54\ 13.37  ,-10\ 24\ 46.4&   0.076&   1.38{\rm  E+44}&  0.95&2.75{\rm E+12}&  2.08{\rm E+12} \\
{\mathrm {CL}}114&   23\ 37\ 40.56  ,+00\ 16\ 36.5&   0.278&   6.30{\rm  E+44}&  1.21&5.16{\rm E+12}&  4.28{\rm E+12} \\
                        \noalign{\smallskip}			    
            \hline					    
            \noalign{\smallskip}			    
            \hline					    
         \end{array}
     $$ 
%\begin{list}{}{}  
%\item[$^{\mathrm{a}}$] Considered among ``best FGs'' following the S07 suggestions; the remaining FGs do not enter in our definition of ``best FGs''.
%\item[$^{\mathrm{b}}$] Considered among ``best FGs'' following the preliminar analysis of FOGO data.
%\item[$^{\mathrm{c}}$] Already analyzed through FOGO data, but not entering in our definition of ``best FGs''. We list available properties.
%\item[$^{\mathrm{d}}$] FG19 was not considered in our analysis being not well covered by SDSS-DR7 data.
%\end{list}
\end{table*}
%\end{table}

%% file: tabfit.tex
%\documentclass[referee]{aa}
%\usepackage{graphicx}
%%new commands
%\def\lesssim{\mathrel{\hbox{\rlap{\hbox{\lower4pt\hbox{$\sim$}}}\hbox{$<$}}}}
%\def\gtrsim{\mathrel{\hbox{\rlap{\hbox{\lower4pt\hbox{$\sim$}}}\hbox{$>$}}}}
%\newcommand{\mincir}{\raise -2.truept\hbox{\rlap{\hbox{$\sim$}}\raise5.truept
%\hbox{$<$}\ }}
%\newcommand{\magcir}{\raise -2.truept\hbox{\rlap{\hbox{$\sim$}}\raise5.truept
%\hbox{$>$}\ }}
%\newcommand{\siml}{\raise -2.truept\hbox{\rlap{\hbox{$\sim$}}\raise5.truept
%\hbox{$<$}\ }}
%\newcommand{\simg}{\raise -2.truept\hbox{\rlap{\hbox{$\sim$}}\raise5.truept
%\hbox{$>$}\ }}
%\newcommand{\be}{\begin{equation}}
%\newcommand{\ee}{\end{equation}}
%\newcommand{\ba}{\begin{eqnarray}}
%\newcommand{\ea}{\end{eqnarray}}
%\newcommand {\h} {$h^{-1}$ Mpc $ \;$}
%\newcommand {\kpc} {$h^{-1}$ kpc}
%\newcommand {\hh} {$h^{-1}$ Mpc}
%\newcommand {\ks} {km~s$^{-1} \;$}
%\newcommand {\kss} {km~s$^{-1}$}
%\newcommand {\mpc} {$Mpc \;$}
%\newcommand {\msun} {$h^{-1} \  M_{\odot} \;$}
%\newcommand {\m} {$M_{\odot} \;$}
%\newcommand {\ml} {$h \, M_{\odot}/L_{\odot} \;$}
%\newcommand {\mll} {$h \, M_{\odot}/L_{\odot}$}
%\newcommand{\vel}{\,{\rm km\,s^{-1}}}
%\newcommand{\tng}{\mathrm{T}}
%\newcommand{\sds}{\mathrm{S}}
%\newcommand{\tns}{\mathrm{T+S}}
%%
%\begin{document}

%\addtocounter{table}{-2}
\begin{table}[!ht]
        \caption[]{Fit parameters obtained using Eq.~5.}
         \label{tabfit}
              $$ 
        % \begin{array}{p{0.5\linewidth}l}
           \begin{array}{l r c c}
            \hline
            \noalign{\smallskip}
            \hline
            \noalign{\smallskip}
%&&&\multicolumn{4}{c}{{\rm Fgs}} & \multicolumn{4}{c}{{\rm RASS-SDSS\ clusters}} \\
Sample& N & a & b\\
           \hline
            \noalign{\smallskip}  
{\rm CONF-FGS}  &12  &-0.3\pm0.1  &1.8\pm0.3\\
{\rm NOCONF-FGS}&16&-0.2\pm0.1  &2.1\pm0.4\\
{\rm ALL-FGS}   &28   &-0.24\pm0.08&2.0\pm0.2\\
{\rm CLs}       &102  &-0.32\pm0.04&1.78\pm0.08\\
\hline
            \noalign{\smallskip}			    
            \hline					    
         \end{array}
     $$ 
\end{table}

%%
%\end{document}